\documentclass{aa}  
\usepackage{graphicx}
\usepackage{txfonts}
\usepackage{natbib}
\usepackage[normalem]{ulem}

\begin{document} 

   \title{The Galah survey and symbiotic stars}
   \subtitle{II. Discovery, validation, and intensive follow-up 
             of 44 new systems accreting at the lowest detectable levels}

   \titlerunning{Galah Symbiotic Stars - II}

   \authorrunning{U. Munari et al.}

   \author{Ulisse~Munari
          \inst{1},
          Paolo~Valisa
          \inst{1,2},
          Nicola~Masetti
          \inst{3,4},
          Massimo~Fiorucci
          \inst{1,2},
          Manica~Perko
          \inst{5},
          Bor~Jamnik
          \inst{5},
          Gregor~Traven
          \inst{5},
          Andrea Frigo
          \inst{2},
          Joss~Bland-Hawthorn           
          \inst{6},
          Sven~Buder                    
          \inst{7},
          Andrew~R.~Casey               
          \inst{8,9},
          Gayandhi~M.~De~Silva          
          \inst{10,11},
          Pradosh B. Das
          \inst{12},
          Ken~C.~Freeman                
          \inst{7},
          Michael~R.~Hayden             
          \inst{6},
          Janez~Kos                     
          \inst{5},
          Geraint~F.~Lewis              
          \inst{6},
          Jane~Lin                      
          \inst{7},
          Sarah~L.~Martell              
          \inst{13},
          Katharine~J.~Schlesinger      
          \inst{7},
          Sanjib~Sharma                 
          \inst{14},
          Jeffrey~D.~Simpson            
          \inst{13},
          Dennis Stello
          \inst{6,13},
          Daniel~B.~Zucker              
          \inst{12,15}, and
          Toma\v{z}~Zwitter             
          \inst{5}
                }

   \institute{  INAF Osservatorio Astronomico di Padova, 36012 Asiago (VI), Italy,\\   
              \email{ulisse.munari@inaf.it}
        \and    
                ANS Collaboration, c/o Astronomical Observatory, 36012 Asiago (VI), Italy
        \and         
                INAF-Osservatorio di Astrofisica e Scienza dello Spazio, via Gobetti 101, 40129 Bologna, Italy 
        \and    
                Departamento de Ciencias F\'isicas, Universidad Andr\'es Bello, Fern\'andez Concha 700, Las Condes, Santiago, Chile 
        \and    
                Faculty of Mathematics and Physics, University of Ljubljana, Jadranska 19, 1000 Ljubljana, Slovenia 
        \and    
                Sydney Institute for Astronomy, School of Physics, A28, The University of Sydney, NSW 2006, Australia 
        \and    
                Research School of Astronomy \& Astrophysics, Australian National University, ACT 2611, Australia 
        \and    
                Monash Centre for Astrophysics, Monash University, Australia 
        \and    
                School of Physics and Astronomy, Monash University, Australia 
        \and    
                Australian Astronomical Optics, Faculty of Science and Engineering, Macquarie Univ., Macquarie Park, NSW 2113, Australia 
        \and    
                Macquarie University Research Centre for Astronomy, Astrophysics \& Astrophotonics, Sydney, NSW 2109, Australia 
        \and    
                School of Mathematical and Physical Sciences, Macquarie University, Sydney, NSW 2109, Australia 
        \and    
                School of Physics, UNSW, Sydney, NSW 2052, Australia 
        \and    
                Space Telescope Science Institute, Baltimore, USA
        \and    
                Macquarie University Research Centre for Astrophysics and Space Technologies, Sydney, NSW 2109, Australia 
             }

   \titlerunning{Galah survey and symbiotic stars II}   

   \date{Received September 15, 1996; accepted March 16, 1997}

  \abstract 
  {
  We report the discovery, validation, and characterization of 44 new
  symbiotic stars (SySts) accreting at the lowest detectable levels (L$_{\rm
  acc}$$\lesssim$3~L$_\odot$).  Candidates were initially selected from the
  Galah survey among non-pulsating M-type red giants exhibiting H$\alpha$
  excess.  Between 2021 and 2025, these targets were subjected to intensive
  multi-wavelength campaigns, including ground-based, high-resolution
  Echelle spectroscopy over 544 nights, multi-epoch flickering searches down
  to 1~mmag, and UV/X-ray observations with the \textit{Swift} satellite. 
  Our high-resolution spectra allowed for a detailed investigation of
  H$\alpha$ flux and line-profile variability.  Crucially, the radial
  velocities measured for the M-giant components yielded reliable
  spectroscopic orbits for 22 objects and preliminary orbits for an
  additional eight, corroborated by binary-induced modulations in ASAS-SN
  photometry.  Measured accretion levels vary by factors of up to
  20$\times$, emphasizing the need for high-sensitivity, long-term
  monitoring.  \textit{Swift} X-ray data exclude neutron star companions in
  all targets, consistent with predominance of white-dwarf degenerates in
  known SySts.  Combined with the 33 objects from Paper~I, this sample of 77
  new SySts significantly exceeds the number of previously known,
  accreting-only Galactic symbiotic stars, constituting the bulk of the
  low-accretion (LowAcc) variety discovered to date and representing $>$10\%
  of all known SySts in the Galaxy.  As SySts likely spend most of their
  lifetimes in LowAcc faint states, characterizing this population is
  essential for estimating the total Galactic SySt population and evaluating
  the role of SySts as Type~Ia supernova progenitors or sources of UV excess
  in elliptical galaxies.
  }

   \keywords{binaries: symbiotic  ---  stars: white dwarfs  --- 
             stars: late-type --- stars: variables: general ---
             techniques: spectroscopic --- surveys}

   \maketitle
   \nolinenumbers

\section{Introduction}

Symbiotic stars (SySts) are interacting binaries composed of a red giant
(RG) and a degenerate companion (usually a white dwarf (WD), or very rarely
a neutron star), typically orbiting with periods of a few years and
separations of a few astronomical units.  The companion accretes from the RG
via wind interception or Roche-lobe overflow; so far, there is little
evidence that magnetic fields play a significant role in this process,
except perhaps in a few isolated cases \citep{1999ApJ...517..919S,
2023A&A...675A.140M, 2026MNRAS.546ag299P}.

The classification criteria for SySts have widened over time.  The original
definition formulated by \citet{1984PASA....5..369A} tended to isolate SySts
undergoing nuclear burning at the surface of a white dwarf
\citep[e.g.,][]{2001MNRAS.326..553S, 2016acps.confE..21S,
2003ASPC..303....9M, 2019arXiv190901389M, 2025Galax..13...49M}.  Later,
\citet{2000A&AS..146..407B} relaxed these constraints to include systems
that are not burning but are accreting onto a WD at a high rate.  The
discovery that some SySts contain a neutron star (NS) instead of a WD
\citep[e.g.,][]{2007A&A...470..331M}---along with the recognition of SU Lyn
as the prototype of a large, previously unsuspected population of SySts
accreting on WDs at (very) low rates \citep{2016MNRAS.461L...1M}---has
shaped the modern definition; i.e., any binary where a WD or an NS accretes
enough material from an RG companion such that the interaction is detectable
at some wavelength is considered a SySt.

The traditional view \citep[e.g.,][]{1986syst.book.....K} implicitly assumed
a steady-state evolution for the different types of SySts; thus, a burning
system was thought to keep burning for its entire symbiotic lifetime.  This
posed the problem of refueling the burning shell, which was circumvented by
assuming that---regardless of radiation pressure or stellar
winds---accretion onto the WD proceeds unabated, matching precisely the rate
at which fuel is consumed.  In a more modern approach \citep[as discussed
by][]{2019arXiv190901389M}, SySts are seen to alternate between burning
phases and accreting-only states; longer intervals are spent accreting
material onto the surface of the WD, which is subsequently ignited once the
conditions for nuclear burning are reached.  When the shell fuel is
exhausted, burning ceases and accretion resumes to prepare for a new cycle.

In addition to being fascinating astrophysical laboratories in their own
right, SySts have far-reaching implications for a broad range of topics. 
These include the identification of progenitors for Type Ia supernovae
\citep[e.g.,][]{1973ApJ...186.1007W, 1992ApJ...397L..87M,
2011ApJ...735L..31C, 2020ApJ...895...70S, 2025A&ARv..33....1R}, the agents
responsible for the ultraviolet (UV) excess in elliptical galaxies
\citep[e.g.,][]{1969PASP...81..475C, 1990ApJ...364...35G,
1999ARA&A..37..603O, 2011Ap&SS.335..193C}, the complex mechanisms shaping
planetary nebulae \citep[e.g.,][]{2011MNRAS.410.1349C, 2017NatAs...1E.117J,
2025IAUS..384.....D}, and the production--destruction cycle for dust grains under hostile
conditions \citep[e.g.,][]{1987PASP...99..573W, 1990MNRAS.242..653M,
2004A&A...417..695S, 2011A&A...530A..44J}.

The relevance of the symbiotic channel scales with the total number of SySts
in the Galaxy, N(SySts), which has been variously estimated in the past
\citep[][among others]{1984PASA....5..369A, 1986syst.book.....K,
1992ApJ...397L..87M, 1993ApJ...407L..81K, 1995ApJ...447..656Y,
2003ASPC..303..539M, 2006MNRAS.372.1389L, 2025A&A...698A.155L}, usually by
counting known objects within a given volume and scaling the results to the
whole Galaxy.  These estimates range over 1,500 $\lessapprox$ N(SySts)
$\lessapprox$ 500,000, suggesting that we still lack a definitive consensus.

These estimates of N(SySts) rested primarily on the few hundred objects
listed in the available catalogs of symbiotic stars
\citep{1984PASA....5..369A, 2000A&AS..146..407B, 2019RNAAS...3...28M,
2019ApJS..240...21A}, which were dominated by burning systems.  Due to the
immense luminosity radiated by the burning shell---even exceeding that of
the RG---these systems are relatively easy to identify throughout the
Galaxy \citep{1984PASA....5..369A} and even within the Local Group
\citep{2003ASPC..303..539M}.  Far more challenging to detect are the
accreting-only SySts, in which accretion processes radiate orders of
magnitude less than a burning shell or the RG.  As outlined in
Appendix~\ref{lifetime}, basic evolutionary considerations suggest that
SySts spend the (vast) majority of their lifetime in an accreting-only state. 
Consequently, this subclass likely represents the dominant population for
the true N(SySts).

In addition, accretion onto the WD during these non-burning intervals may
vary by orders of magnitude between high-accretion (HighAcc) and low-accretion (LowAcc) states,
as exemplified by T CrB, AT 2023txn, and SU Lyn \citep{2018ApJ...864...46L,
2020ApJ...902L..14L, 2021MNRAS.505.6121M, 2025A&A...701A.176M,
2025A&A...702A.269M}, which spend most of their time in LowAcc states
characterized by a weak UV excess, feeble emission lines (usually detectable
only in high-resolution spectra), and accretion luminosities of $L_{\rm
acc}$$\sim$10$^{0-1}$~L$_\odot$.  At least in T~CrB, these low states seem
to be corresponding to the gradual buildup of mass in the accretion disk, with
little material actually transferred to the central WD.  Only when the disk
grows beyond stable conditions does it flush most of its content onto the WD
in a bright, shorter lived HighAcc episode ($L_{\rm
acc}$$\sim$10$^2$~L$_\odot$) marked by strong UV excess and intense
emission lines. If this behavior is common, the majority of accreting-only
SySts in the Galaxy are expected to be in a LowAcc state.

In this series of papers, we aim to discover SySts precisely in the LowAcc
state.  Although they are the most difficult to identify, they represent the
most abundant population in the Galaxy and are therefore the most reliable
tracers to constrain N(SySts).  To this end, we targeted non-pulsating RGs
identified in the high-resolution spectra collected by the Galah survey of
the southern sky \citep{2015MNRAS.449.2604D, 2017MNRAS.464.1259K,
2021MNRAS.506..150B, 2025PASA...42...51B} that exhibit even minimal
H$\alpha$ emission.  We subjected these candidates to extensive
multiwavelength and multi-epoch follow-up observations to confirm their
nature as binaries with a degenerate accreting companion.  The Galah survey
provides a representative and statistically well-characterized sample of
Galactic RGs.  This enables a robust estimate of the symbiotic to any other
RG ratio, which can be extrapolated to the entire Galaxy while minimizing
the systematic uncertainties associated with the modeling of the population of
binaries. In the course of our analysis of the spectra collected by
the Galah survey, we also identified SySts in the HighAcc
regime as well as burning-type systems (some of which were previously
known).  These sources will be discussed in a separate paper, as the present
work focused solely on SySts in a LowAcc state.

Following an initial sample of 33 objects studied in \citet[][hereafter
Paper I]{2021MNRAS.505.6121M}, this Paper II identifies a further 44
symbiotic stars accreting at the lowest detectable levels (a few L$_\odot$). 
These 77 accreting-only symbiotic stars already outnumber previously known
cases, and the investigation of an additional large sample is well underway
for the forthcoming Paper III.  Recent years have witnessed a surge in the
search for new SySts, as reviewed in Appendix~\ref{appendix:literature}. All
newly discovered SySts are almost invariably of the burning type or the
HighAcc variant of the accreting-only type.  To date, no other search has
reached our level of sensitivity toward the LowAcc variant or provided the
extensive multiwavelength follow-up observations (spanning several years)
required to characterize the main properties of these new SySt systems.

\section{Target selection}\label{target}

The target selection for the present study was conducted in 2019,
concurrently with that for Paper I, where more comprehensive details and
quantitative definitions are provided.  Only a brief summary is presented
here.  We focused on M-type red giants, as they are the most abundant among
known SySts. Systems containing other types of cool giants
will be discussed in subsequent papers of this series.

Given the high galactic latitude (and consequently low reddening) of the
Galah observations, red giants were initially selected by applying the
following criteria: ($i$) 2MASS color $(J-K_s)$$\geq$0.90, and ($ii$) Gaia
parallax $\pi$$\leq$3 mas.  To isolate those with an M-type spectrum, we
inspected the Galah spectra for TiO absorption bands, in the process also picking
up some S-type giants whose spectra exhibit both TiO and molecular
bands associated with s-process elements (specifically ZrO).  The M giants
were then organized into the temperature sequence defined in Paper~I,
consisting of 13 $f$-bins based on the depth of the TiO band-heads.  The
boundaries of these bins were defined to ensure a comparable number of
objects (several thousands) in each, at least for the lower $f$-bins (hotter
M giants), where the majority of the population resides.  Median template
spectra were subsequently derived for each $f$-bin.  A calibration of the
$f$-bins sequence in terms of absolute magnitudes and intrinsic colors is
provided in Appendix~\ref{appendix:absmag}, where we also discuss the
shortcomings of the classical M0--M10 III classification scheme when 
applied to high-resolution spectra and their quantitative analysis.

The subsequent step involved searching the Galah spectra of these M giants
for H$\alpha$ and H$\beta$ emission signatures.  To achieve this, the
appropriate templates were subtracted from the continuum-normalized spectra,
and the residuals were inspected for H$\alpha$ emission.  The resulting
candidates were then all visually verified to eliminate false positives. 
While Paper I focused on M giants with an H$\alpha$ emission profile
exceeding a threshold of 0.5 in the subtracted spectrum, the present work
extends this selection to a lower threshold of 0.1.

Not all objects exhibiting H$\alpha$ emission are necessarily valid
candidates for accreting-only SySts.  The primary sources of false positives
are radial pulsators (e.g., Miras or SRa variables), which trigger Balmer
line emission via shocks forming in the deep atmosphere during the pulsation
cycle.  We filtered out Miras and SR variables by discarding objects whose
ASAS-SN light curves are modulated by signatures of radial pulsation with an
amplitude $\geq$0.5 mag.  As a secondary pruning criterion, we examined the
relative intensities of the H$\beta$ and H$\alpha$ emission in the
subtracted spectra.  Since radially pulsating stars invariably show
H$\beta$/H$\alpha > 1$ on subtracted spectra, we only retained systems where
H$\beta$/H$\alpha \ll 1$.

It is, of course, possible for a Mira to be paired with a degenerate companion
in a SySt; $o$ Cet itself, the prototype of Mira variables, is a well-known,
accreting-only SySt.  Indeed, Hubble (UV) and Chandra (X-rays) have
spatially resolved its WD companion, the accretion disk, and the
fueling stream \citep{2006ESASP.604..183K}.  Candidate Galah SySts
containing pulsating variables will be addressed elsewhere in this series.

The $\sim$100 final candidates selected for the present Paper II were all
located at $\delta \geq -27^\circ$ to ensure accessibility from the
telescopes in Asiago and Varese.  In 2021, we began collecting
high-resolution Echelle spectra and conducting photometric flickering runs,
supplemented by UV and X-ray observations obtained with the {\it Swift}
satellite.  After two years, we shortlisted 60 objects selected for being
the most promising and/or having the most extensive follow-up data.  These
were monitored intensively for an additional three years (2023--2025),
ultimately leading to the discovery, validation, and characterization (over
a total five-year period) of the 44 new and confirmed LowAcc SySts
presented in this work and listed in Table~\ref{tab:newsymb}.
The confirmatory and follow-up observations (2021–2025) for the new SySts
are described in Sects.~\ref{Echelle}--\ref{Swift}, with the bulk of
the technical specifications provided in
Appendices~\ref{appendix:Echelle}--\ref{appendix:Swift}.

\section{Radial velocities, orbits, and H$\alpha$ emission from
high-resolution Echelle spectroscopy}\label{Echelle}

\begin{table} 
\caption{Log-book summarizing the 591 high-resolution spectroscopic
observations (one per night) that we obtained for the 44 new symbiotic
stars, along with the red giants' heliocentric radial velocities derived via
cross-correlation.}
\begin{tabular}{@{}c@{~~~}c@{~~~}c@{~~~}r@{~~}r@{~~}r@{}}
\hline\hline
&&&\\
     &      &     & \multicolumn{3}{c}{Heliocentric RV}\\ \cline{4-6}
Name & date & HJD & Varese & Asiago & Galah\\
&&&\\
\hline
&&&\\
GaSS 2-69 & 2024-08-20 & 2460543.37  &          & $-$125.5    &           \\
GaSS 2-69 & 2025-07-18 & 2460875.42  &$-$126.8  &             &           \\
GaSS 2-69 & 2025-08-07 & 2460895.42  &          & $-$125.3    &           \\
GaSS 2-69 & 2025-09-17 & 2460936.40  &$-$123.9  &             &           \\
GaSS 2-69 & 2025-10-17 & 2460966.36  &$-$123.4  &             &           \\
GaSS 2-70 & 2015-06-06 & 2457180.11  &          &             &  107.8    \\
GaSS 2-70 & 2021-08-20 & 2459447.45  &          &    110.8    &           \\
GaSS 2-70 & 2024-08-08 & 2460531.37  &   107.1  &             &           \\
GaSS 2-70 & 2025-08-09 & 2460899.42  &          &    111.7    &           \\
&&&\\
\hline 
\end{tabular}   
\tablefoot{The full table is only available electronically via CDS, a small
portion is shown here for guidance on its form and content.  The RVs are
expressed in km\,s$^{-1}$.}
\label{tab:logEchelle}
\end{table} 

A large number of high-resolution optical spectra of the new 44 SySts were
collected between 2021 and 2025 using the Echelle spectrographs mounted on
the Asiago 1.82m and Varese 0.84m telescopes.  A log of these observations
is provided in Table~\ref{tab:logEchelle}.  In total, Echelle spectra were
obtained over 544 distinct nights, with an average of 12.5 visits per
program star.  The Galah survey provided the initial spectrum for each
object (with a mean date of June 2016), contributing a further 47 epochs
(one per object, plus a second visit in some cases).  This brings the grand
total to 591 visits, or an average of 13.5 per object.  Details of the
telescopes, spectrographs, and observing and data reduction procedures are
provided in Appendix~\ref{appendix:Echelle}.

\subsection{Radial velocities}\label{radvel}

The radial velocities (RVs) were measured on the single, one-dimensional,
continuum-normalized version of the Echelle spectra via Fourier
cross-correlation against suitable templates \citep{1979AJ.....84.1511T}, as
implemented by {\sc rv/fxcor} in IRAF.  To maximize the signal-to-noise
ratio (S/N) of the object spectra and to avoid regions affected by telluric
lines, interstellar NaI, and possible stellar emission lines, the
cross-correlation was restricted to three spectral intervals: 4900--5800,
6000--6250, and 6330--6450~\AA. The list of measured RVs is provided 
in Table~\ref{tab:logEchelle}.

The spectra collected with the Asiago Echelle spectrograph were all
cross-correlated against the same template, selected from the synthetic
spectral library of \citet{2005A&A...442.1127M}.  The template was chosen
for a resolving power of 20,000, with atmospheric parameters of T$_{\rm
eff}$=3500~K, $\log g$=1.5, [Fe/H]=0.0, [$\alpha$/H]=0.0, and
$\zeta$=2~km\,s$^{-1}$.  This approach has already proven to be very
effective in obtaining the best available RV orbit for the M3\,III red giant
in the symbiotic recurrent nova T~CrB from long-term Asiago and SMARTS-Chiron
Echelle observations \citep{2025A&A...701A.176M}, avoiding systematic
offsets and providing RVs accurate to $\sigma$=0.27~km\,s$^{-1}$ over a
multi-year timescale.

   \begin{figure*}
   \centering
   \includegraphics[width=\textwidth]{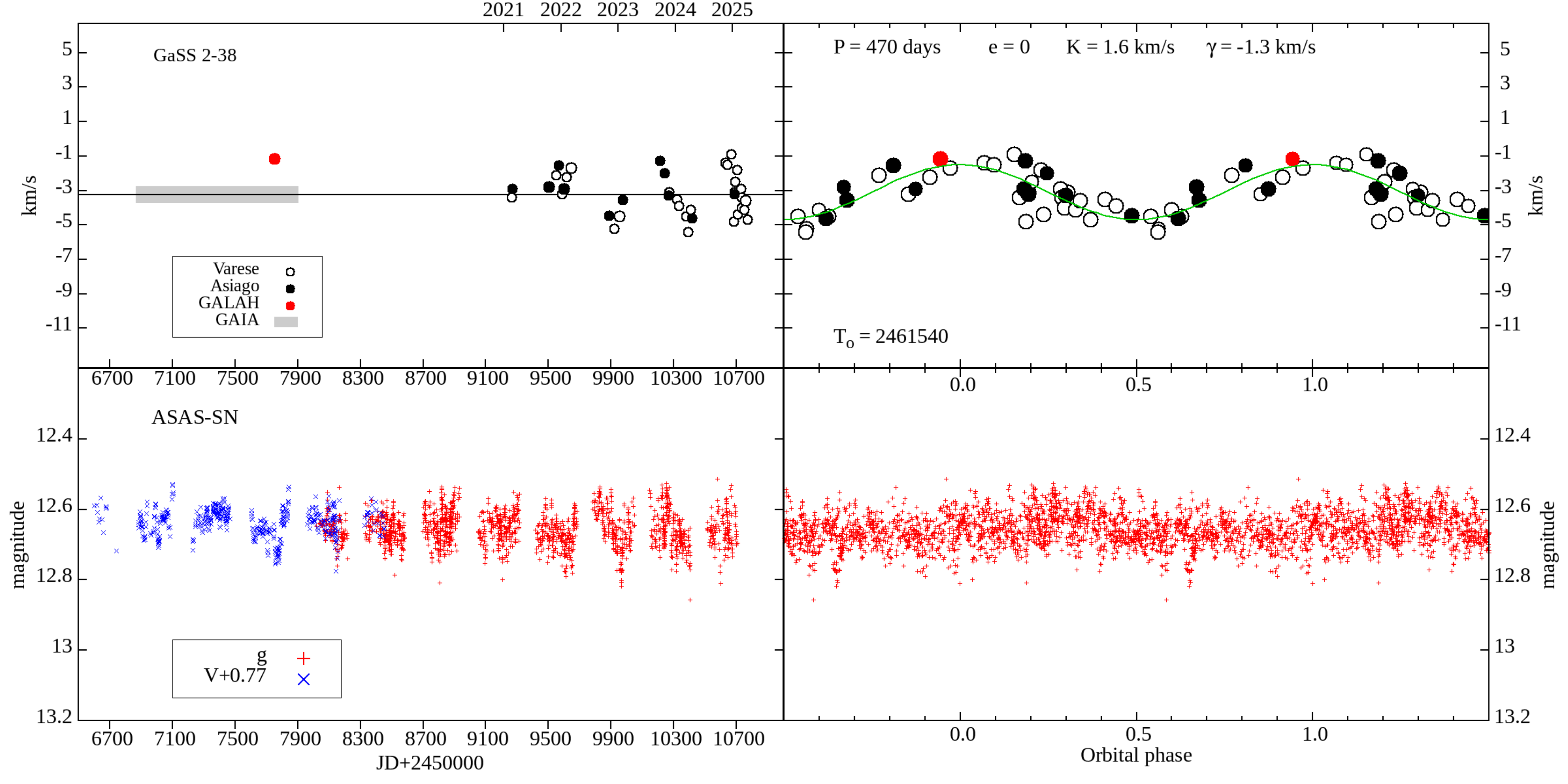}
      \caption{Example of the RVs collected with the
      Asiago 1.82m + Echelle and Varese 0.84m + Echelle telescopes for the
      program stars and the corresponding orbital solution, in this case for
      GaSS 2-38.  Upper left: Solid line shows the Gaia DR3 mean RV, the
      horizontal gray band marks the time covered by Gaia DR3 observations,
      the red dot is the RV from the Galah spectrum, and the open and solid
      black dots mark Varese and Asiago epoch RVs, respectively.  Upper
      right: RVs on the left folded onto the spectroscopic orbit given in
      Table~\ref{tab:orbits} (green sinusoidal line).  Lower left: $V$- and
      $g$-band photometry from ASAS-SN.  Lower right: ASAS-SN photometry in
the      $g$ band folded to the ephemeris of the spectroscopic orbit above.}
   \label{fig:RV}
   \end{figure*}

A different measurement approach was adopted for the spectra collected with
the Varese telescope to enhance its performance to better match that of the
Asiago Echelle spectrograph, which operates at a higher resolving power. 
The Gaia DR3 catalog of radial velocities was searched for a suitable RV
template located within 2$^\circ$ of each program star.  For each
observation of a program star, the corresponding template was observed
immediately before and after.  The object spectrum was then cross-correlated
separately against the two spectra of the template, and the results were
averaged.  The median difference between these two RV measurements is
0.29~km\,s$^{-1}$, a stability comparable to that of the Asiago data.  The
multi-epoch observations of the selected templates were finally mutually
cross-correlated to verify their RV constancy over the 2021--2025 interval. 

\begin{table} 
\caption{Spectroscopic orbital parameters for the red giant components of
the program stars.  }
\begin{tabular}{@{~~}crrrccc@{~~}}
\hline\hline
&&&\\
Name & P & \multicolumn{1}{c}{K} & \multicolumn{1}{c}{$\gamma$} & e & $\omega$ & T$_0$ \\
&&&\\
\hline
&&&\\
&\multicolumn{6}{c}{\it good orbits}\\
&&&\\                                                                          
GaSS 2-34  &     650   &      1.8  &     $-$11.8 &     0  &     &  2461360 \\  
GaSS 2-35  &     380   &      2.7  &     $-$35.0 &     0  &     &  2461480 \\  
GaSS 2-38  &     470   &      1.6  &      $-$1.3 &     0  &     &  2461540 \\  
GaSS 2-40  &     570   &      1.2  &     $-$46.2 &     0  &     &  2461590 \\  
GaSS 2-44  &     342   &      2.3  &     $-$31.5 &   0.2  & 120 &  2461490 \\  
GaSS 2-54  &     335   &      6.3  &     $-$67.0 &   0.2  & 190 &  2461650 \\  
GaSS 2-56  &     690   &      5.7  &    $-$100.8 &     0  &     &  2461590 \\  
GaSS 2-76  &     412   &      2.4  &     $-$86.5 &     0  &     &  2461610 \\  
&&&\\
&\multicolumn{6}{c}{\it fair orbits}\\
&&&\\
GaSS 2-39  &    2000   &      7.6  &        88.9 &     0  &     &  2461580 \\  
GaSS 2-45  &     600   &      1.9  &       133.8 &     0  &     &  2461625 \\  
GaSS 2-51  &     170   &      3.5  &     $-$62.0 &     0  &     &  2461650 \\  
GaSS 2-52  &     840   &     11.0  &     $-$60.0 &     0  &     &  2461640 \\  
GaSS 2-55  &    2250   &      1.6  &        16.0 &     0  &     &  2460340 \\  
GaSS 2-58  &     335   &      3.0  &    $-$106.5 &     0  &     &  2461514 \\  
GaSS 2-60  &     800   &      3.7  &        77.0 &     0  &     &  2461934 \\  
GaSS 2-67  &     450   &      2.0  &     $-$85.5 &     0  &     &  2461720 \\  
GaSS 2-68  &     370   &      2.0  &       116.0 &     0  &     &  2461620 \\  
GaSS 2-69  &     310   &      2.9  &    $-$123.0 &   0.2  & 180 &  2461650 \\  
GaSS 2-71  &     600   &      2.3  &        20.5 &     0  &     &  2461280 \\  
GaSS 2-73  &    3060   &      1.3  &     $-$93.0 &     0  &     &  2460420 \\  
GaSS 2-74  &     116   &      1.7  &        23.5 &     0  &     &  2461050 \\  
GaSS 2-75  &     380   &      2.7  &      $-$7.5 &     0  &     &  2461620 \\  
&&&\\
&\multicolumn{6}{c}{\it tentative orbits}\\
&&&\\
GaSS 2-41  &     220   &      1.7  &        94.5 &     0  &     &  2461530 \\ 
GaSS 2-43  &     249   &      2.4  &     $-$62.5 &     0  &     &  2461480 \\ 
GaSS 2-46  &     720   &      2.0  &       323.5 &     0  &     &  2461870 \\ 
GaSS 2-48  &    4500   &      1.9  &     $-$14.3 &     0  &     &  2461625 \\ 
GaSS 2-63  &     400   &      3.0  &     $-$94.5 &     0  &     &  2461414 \\ 
GaSS 2-64  &    1350   &      4.0  &         7.0 &     0  &     &  2461620 \\ 
GaSS 2-65  &     550   &      1.5  &        62.0 &     0  &     &  2461000 \\ 
GaSS 2-70  &     650   &      2.7  &       109.5 &     0  &     &  2461520 \\ 
&&&\\
\hline 
\end{tabular}   
\tablefoot{The classification into good, fair, and tentative
is explained in Sect.~\ref{spectroscopicorbits}.
The period (P) is expressed in days, the semi-amplitude (K) and
the barycentric velocity ($\gamma$) are in km\,s$^{-1}$, the argument of the
periastron ($\omega$) for eccentric orbits (e$\neq$0) is in degrees, and the
time of passage at ascending quadrature (T$_0$, corresponding to maximum
velocity) is in JD.}
\label{tab:orbits}
\end{table} 

Finally, the Galah radial velocities were imported from Galah Data Release~4
\citep{2025PASA...42...51B}.

\subsection{Spectroscopic orbits}\label{spectroscopicorbits}

The RVs from Table~\ref{tab:logEchelle} were analyzed together with ASAS-SN
$g$-band photometry \citep{2014ApJ...788...48S, 2023arXiv230403791H} to
search for orbital motion.  The orbital period was searched for
simultaneously in the RVs and photometric data using the phase dispersion
minimization algorithm of \citet{1978ApJ...224..953S} and the Fourier
analysis for unequally spaced data of \citet{1975Ap&SS..36..137D}.  To be
accepted as a viable photometric period, the corresponding folded light
curve must display the shape expected from orbital modulation (e.g.,
reflection effect, eclipses, Roche-lobe filling, or transit at the inferior
conjunction of the hot spot) and be properly phased with respect to the
folded RV curve.

For 30 of the 44 program stars, it was possible to obtain an orbital
solution, as reported in Table~\ref{tab:orbits}; for all of these, the
photometric orbital period (where detected) converges to the same value
derived from the radial velocities.  An example of the derived orbits is
provided in Fig.~\ref{fig:RV} for the case of GaSS 2-38\footnote{Equivalent
plots for all other program stars are available at
\texttt{www.astrogeo.va.it/Galah\_Paper2/} or
\texttt{symbiotic-stars.com/$\sim$pub/Galah\_Paper2/}.}. The 470-day period
of GaSS 2-38 is typical for SySts \citep[which usually range between 1 and 3
years;][]{2003ASPC..303....9M}, and the small semi-amplitude
$K$=1.6~km\,s$^{-1}$ suggests the combination of a low orbital inclination
and a low-mass companion (for comparison, a WD of 1.0~M$_\odot$ at
$i$=90$^\circ$ would have caused a semi-amplitude of $K$=14.9~km\,s$^{-1}$
for an RG companion of 1.5~M$_\odot$).  The amplitude of the photometric
modulation of GaSS 2-38 in Fig.~\ref{fig:RV} is $\Delta g$=0.04~mag; its
single maximum, centered around orbital phase 0.25, suggests that its origin
is the side of the RG irradiated by the hot companion.  In contrast, a
single maximum around phase 0.65 would point to the source being the hot
spot transiting at inferior conjunction \citep[see Figure~8
of][]{2025A&A...701A.176M}, while a double maximum centered at phases 0.0
and 0.5 would support an ellipsoidally distorted RG filling its Roche lobe.

The 30 orbital solutions listed in Table~\ref{tab:orbits} were graded using
a quality flag.  Rather than relying on a strict analytical
definition---which would be difficult to formulate given the complex
requirements involving period convergence between photometry and RVs, light
curve morphology and phasing consistent with known SySt variability types,
the number of RV data points, their rms scatter, and their orbital phase
coverage---this flag reflects an overall qualitative assessment of each
orbit.

For those labeled "good", it is unlikely that new observations will
substantially change the orbital parameters (the period in particular).  For
the "tentative" group, the listed orbit is the best that can be derived
given the current data limitations (too few RVs and/or low-quality ASAS-SN
photometry), while "fair" orbits represent an intermediate quality level. 

Quantitatively, for good orbits, the $N$ available RVs yield an average
rms=0.95~km\,s$^{-1}$ and rms\,$/\sqrt{N}$=0.18~km\,s$^{-1}$.  These values
degrade to 1.60 and 0.42~km\,s$^{-1}$ for fair orbits and show a further
increase for tentative ones.

Seven of the remaining 14 program stars show only a photometric period.  For
GaSS 2-37, GaSS 2-59, and GaSS 2-77 (periods of 540, 580, and 650 days,
respectively), the light-curve shape resembles that of a Roche-lobe-filling
RG seen at low orbital inclination, or one only partially filling the lobe
(small photometric amplitude).  For the other four stars in this group,
namely GaSS 2-42, GaSS 2-61, GaSS 2-62, and GaSS 2-66, with periods of 340,
372, 375, and 360 days, respectively, the light curves show single maxima
and minima: given the proximity to the 365-day yearly alias, a photometric
coverage longer than the $\sim$10~years currently available from ASAS-SN would
be necessary to fully cover all phases and confirm the periods.
Finally, for the last seven stars (GaSS 2-36, GaSS 2-47, GaSS 2-49, GaSS
2-50, GaSS 2-53, GaSS 2-57, and GaSS 2-72), the search for a suitable
periodicity in both the photometric and RV data was unsuccessful.

\subsection{H$\alpha$ profiles from subtracted spectra}\label{emissionline}

For our targets, the H$\alpha$ emission is sufficiently weak that properly
detecting the line, extracting a clean profile, and measuring the radiated
flux requires subtracting a template spectrum of the underlying M giant. 
Given the similar resolution of the Asiago and Galah spectra
\citep[e.g.,][]{2023MNRAS.521.3727V}, we adopted the 13 $f$-bin Galah
templates defined by \citet{2021MNRAS.505.6121M}.  We implemented a C++
pipeline based on minimum distance estimation (MDE) to identify the best-fit
$f$-bin template by minimizing the $\chi^2$ between the observed and
template spectra.  The detection and analysis of H$\alpha$ emission profiles
are restricted to the Asiago spectra; the lower resolution of the Varese
observations makes it unreliable to use the same set of Galah templates,
even if appropriately degraded.

   \begin{figure*}
   \includegraphics[width=\textwidth]{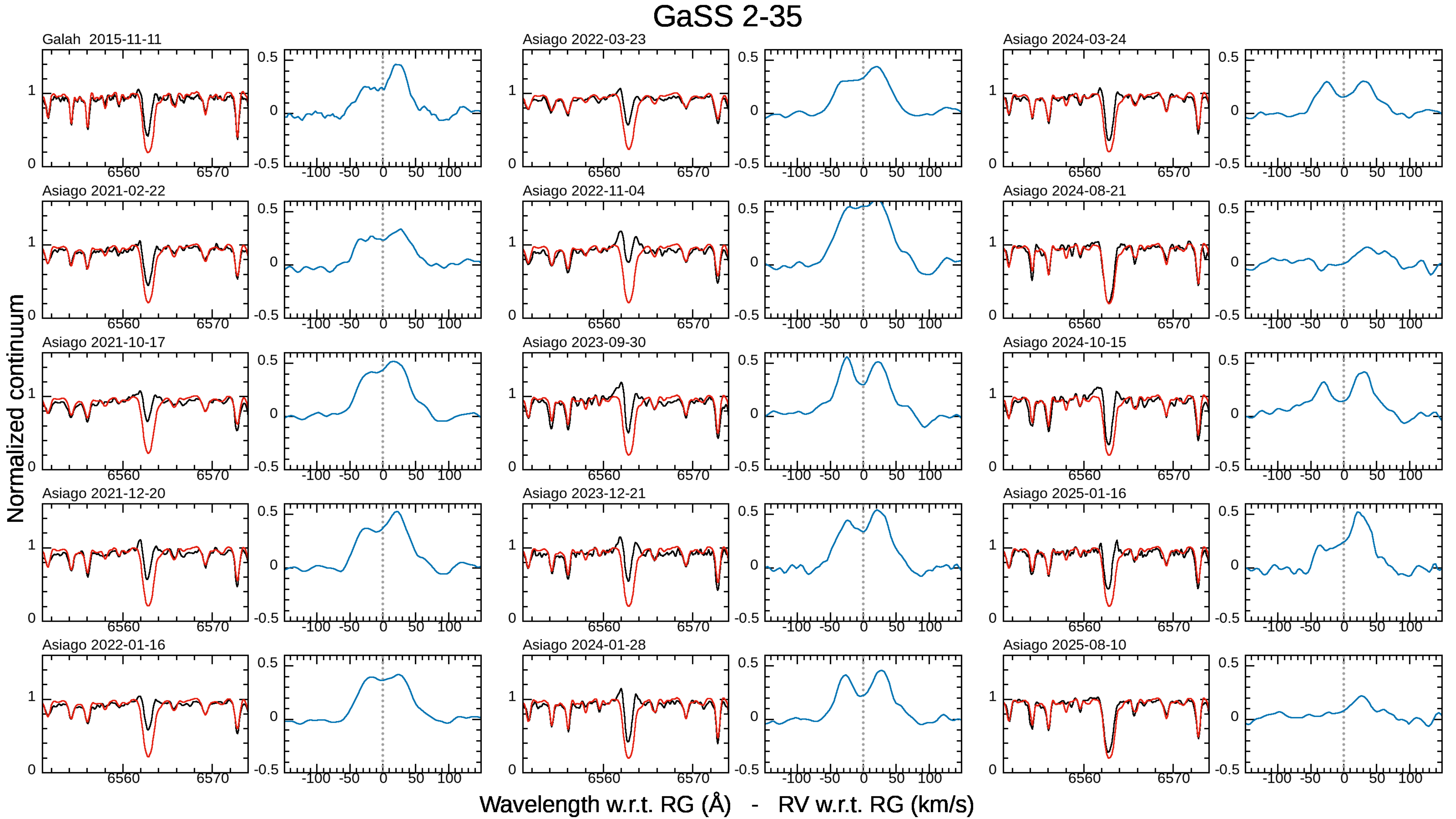}
        \caption{Example of the profile evolution of the H$\alpha$ emission
                 on the high-resolution spectra from the Galah survey and
                 the Asiago 1.82m + Echelle observations, in this case a
                 total of 15 epochs for GaSS 2-35 distributed over
                 2015--2025.  For each observing date (marked at the top
                 left) the observed, continuum-normalized spectrum is
                 plotted in black and that of the corresponding Galah
                 template for the same $f$-bin is drawn in red.  The
                 subtraction of the template from the epoch spectrum
                 produces the H$\alpha$ profile plotted in blue at right.}
        \label{fig:Ha_evolution}
   \end{figure*}

\begin{table*} 
\caption{Properties of the pure-H$\alpha$ emission profile isolated by
subtraction of the red giant from the observed spectrum.  }
\begin{center}
\begin{tabular}{cccc@{~~~}c@{~~~}c@{~~~}c@{~~}c@{~~}c@{~~}c@{~}c@{~~}c@{~~}c@{~~}c}
\hline\hline
&&&\\
          &        &            &\multicolumn{3}{c}{Whole profile}  & &\multicolumn{3}{c}{Emission}&&\multicolumn{3}{c}{Absorption}\\ \cline{4-6} \cline{8-10} \cline{12-14}
Name      & tel.   & date       &  peak & EW  & flux      & & flux      &RV$_{\rm wrt\, RG}$   & FWHM && flux &  RV$_{\rm wrt\, RG}$   &  FWHM\\
          &        &            &       &(\AA)  &           & & &(km/s)  &(km/s) && &(km/s)  &(km/s)\\
&&&\\
\hline
&&&\\
GaSS 2-35 & Galah  & 2015-11-11 & 0.463 & 0.590 & 4.0e-14   & & 1.5e-13   &  +1.0  &   48  &&  1.1e-13 &$-$0.7  &   40 \\ 
GaSS 2-35 & Asiago & 2021-02-22 & 0.338 & 0.557 & 3.7e-14   & & 1.2e-13   &  +1.3  &   55  &&  8.4e-14 &  +0.5  &   45 \\ 
GaSS 2-35 & Asiago & 2021-10-17 & 0.520 & 0.913 & 4.7e-14   & & 6.8e-14   &  +1.3  &   63  &&  2.1e-14 &$-$1.6  &   37 \\ 
GaSS 2-35 & Asiago & 2021-12-20 & 0.526 & 0.817 & 5.0e-14   & & 2.0e-13   &$-$0.7  &   51  &&  1.5e-13 &$-$1.6  &   44 \\ 
GaSS 2-35 & Asiago & 2022-01-16 & 0.420 & 0.757 & 4.6e-14   & & 8.6e-14   &  +2.3  &   57  &&  3.9e-14 &  +1.9  &   39 \\ 
GaSS 2-35 & Asiago & 2022-03-23 & 0.443 & 0.717 & 5.2e-14   & & 2.2e-13   &$-$1.8  &   53  &&  1.7e-13 &$-$2.7  &   46 \\ 
GaSS 2-35 & Asiago & 2022-11-04 & 0.624 & 1.166 & 5.4e-14   & & 1.0e-13   &  +0.0  &   61  &&  4.7e-14 &$-$0.9  &   44 \\ 
GaSS 2-35 & Asiago & 2023-09-30 & 0.562 & 0.961 & 4.9e-14   & & 6.9e-14   &$-$2.4  &   67  &&  2.0e-14 &$-$2.5  &   29 \\ 
GaSS 2-35 & Asiago & 2023-12-21 & 0.542 & 0.883 & 5.4e-14   & & 1.7e-13   &$-$0.2  &   54  &&  1.1e-13 &$-$1.1  &   43 \\ 
GaSS 2-35 & Asiago & 2024-01-28 & 0.459 & 0.738 & 4.5e-14   & & 1.4e-13   &  +0.2  &   55  &&  9.2e-14 &$-$0.6  &   41 \\ 
GaSS 2-35 & Asiago & 2024-03-24 & 0.305 & 0.533 & 3.9e-14   & & 1.2e-13   &  +2.5  &   58  &&  8.1e-14 &  +2.1  &   44 \\ 
&&&\\
\hline 
\end{tabular}   
\tablefoot{The full table is only available electronically via CDS; a small
portion is shown here for guidance on its form and content.  "Emission" and
"Absortion" refer to fitting the "Whole profile" (blue spectrum in
Fig.~\ref{fig:Ha_evolution}) with a combination of two Gaussians, one in
emission and the other in absorption.  The fluxes are in
erg\,cm$^{-2}$\,s$^{-1}$, and the radial velocities are given with respect
to the red giant.}
\end{center}
\label{tab:Ha_evolution}
\end{table*} 

The procedure begins with initial data cleaning for cosmic rays and telluric
lines, followed by a preliminary wavelength alignment.  The observed
spectrum is then locally normalized using a moving average and compared to
the templates.  The MDE algorithm iteratively optimizes both the
radial-velocity shift and the Gaussian broadening required to match the
instrumental resolution.  During this process, the cores of the emission
lines (e.g., 6561.3--6564.3~\AA\ for H$\alpha$) are masked.  After
fine-tuning the normalization and resolution, the optimal $f$-bin template
was identified.  The S/N and instrumental stability are high enough that the
optimal $f$-bin template remains constant for all the observations of a
given object.  Finally, the template was subtracted, leaving a pure emission
residual for further analysis (the profile in blue in
Fig.~\ref{fig:Ha_evolution}).

An example of the results is presented in Fig.~\ref{fig:Ha_evolution} for
the program star GaSS~2-35.  The 15 observing epochs, spanning from 2015 to
2025, highlight the variability of H$\alpha$ in both profile shape and
radiated flux.  Similar figures illustrating the evolution of the H$\alpha$
profile for the remaining program stars are available from the same web site
as given in footnote~1.  The basic parameters extracted from these cleaned
H$\alpha$ profiles are summarized in Table~\ref{tab:Ha_evolution}, along
with the results of fitting these profiles with a combination of two
Gaussians (one for the emission and one for the absorption component).  We
also experimented with other profile types (Lorentzian, Voigt) but found no
systematic statistical improvement over Gaussian fits.  The H$\alpha$
profile deconvolution reported in Table~\ref{tab:Ha_evolution} is intended
purely to provide dataset-wide inputs for exploring the potential orbital
modulations in the next section; individual fits are not intended as
high-precision physical models of single profiles that would warrant formal
error propagation or their explicit quoting in Table~\ref{tab:Ha_evolution}.

\subsection{Orbital modulation of H$\alpha$ emission}\label{Hamodulation}

To spatially locate the source of H$\alpha$ emission in the 44 newly
discovered LowAcc SySts, we investigated the properties of the pure
H$\alpha$ emission profiles isolated by subtracting the red giant
contribution from the observed spectra.  We focused on the 22 objects with
"good" or "fair" orbital solutions (Table~\ref{tab:orbits}).
We explored the relationship between orbital phase and the H$\alpha$
parameters listed in Table~\ref{tab:Ha_evolution}, but found no clear
pattern either on an individual object basis or when phase-averaging all
systems together.

The lack of clear repeatability with orbital phase for individual objects
suggests that stochastic accretion variability dominates over the effects of
orbital geometry.  Furthermore, the absence of a distinct trend in the
phase-averaged ensemble suggests a combination of accretion modes, likely
involving both wind-capture and Roche-lobe overflow.

   \begin{figure*}
   \centering
   \includegraphics[width=16.8cm]{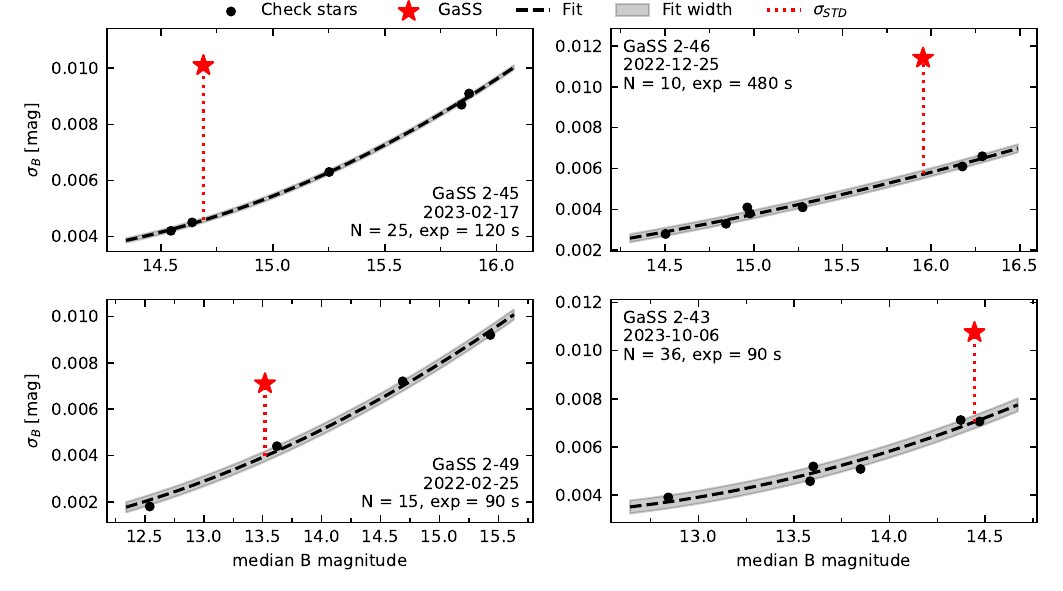}
      \caption{Examples of the results obtained in the search for flickering
               in the $B$ band with the Asiago 67/92 cm Schmidt telescope
               (see Table~\ref{tab:flickering}, Sect.~\ref{flickering}, and
               Appendix~\ref{appendix:flickering} for details).  The
               dispersion $\sigma$ of the $N$ consecutive measurements of
               duration {\it exp} for the GaSS symbiotic star (red symbol)
               is compared to that of the check stars of a similar red color
               (gray band).}
         \label{fig:flickering}
   \end{figure*}

\section{Flickering observations}\label{flickering}

The flickering observations for Paper~I were acquired with a 50cm telescope,
featuring a 40 arcmin, well-corrected field of view operated robotically from
San Pedro de Atacama, Chile.  For the present Paper~II, all flickering
observations were carried out robotically with the Asiago 67/92cm
Schmidt telescope (focal length 2.15m).  The telescope is equipped with a
Moravian CCD camera (KAF-16803 detector, 4096$\times$4096 active matrix, 9
$\mu$m pixel size), covering a field of 59$\times$59 arcmin at 0.86
arcsec/pix.

This setup offers several advantages compared to Paper~I: (1) a
$\sim$2$\times$ gain in light-collecting power, significantly benefiting the
S/N and enabling the exploration of shorter flickering time-scales; (2) a
field of view twice as large (perfectly corrected by a 3-element
field-flattener), allowing for a broader choice of both the photometric
comparison sequence and check stars; (3) tighter control over real-time
focusing and guiding (fractional-pixel stability over hours) to fully
exploit the excellent seeing conditions (typically FWHM=1.1 arcsec); (4)
higher overall observational accuracy, reducing the dispersion on check
stars to as little as 0.002 mag over one-hour runs; and, in particular, (5)
the possibility of multiple visits to each target.

\begin{table*} 
\caption{Results of multi-epoch visits to the program stars in search for
flickering in the $B$ band.}
\begin{tabular}{cccccccccrl}
\hline\hline
\multicolumn{1}{c}{(1)}&\multicolumn{1}{c}{(2)}&\multicolumn{1}{c}{(3)}&\multicolumn{1}{c}{(4)}&\multicolumn{1}{c}{(5)}&\multicolumn{1}{c}{(6)}&\multicolumn{1}{c}{(7)}&\multicolumn{1}{c}{(8)}&\multicolumn{1}{c}{(9)}&\multicolumn{1}{c}{(10)}&\multicolumn{1}{c}{(11)}\\
&&&\\
Name & date & HJD & expt & N($B$) & $B$    & $\sigma$($B$) & diff. & RMSE & $\sigma_{\rm STD}$ & notes \\
     &      &     &(sec) &      & (mag)&  (mag)      & (mag) & (mag)&                   &       \\
&&&\\
\hline
&&&\\
GaSS 2-45 & 2022-02-25 & 2459635.626 & 120 &  25 & 15.089 & 0.0081  & 0.0013 & 0.0007 &   1.9 &     \\
GaSS 2-45 & 2023-02-17 & 2459992.524 & 120 &  25 & 14.690 & 0.0101  & 0.0055 & 0.0001 &  70.4 &     \\
GaSS 2-45 & 2023-11-12 & 2460260.596 & 130 &  17 & 15.233 & 0.0126  & 0.0039 & 0.0014 &   2.8 &     \\
GaSS 2-46 & 2022-12-25 & 2459938.678 & 480 &  10 & 15.954 & 0.0114  & 0.0057 & 0.0002 &  28.7 &     \\
GaSS 2-46 & 2023-02-17 & 2459992.568 & 480 &  10 & 16.075 & 0.0042  & 0.0003 & 0.0006 &   0.5 & b   \\
GaSS 2-47 & 2022-02-24 & 2459634.614 &  30 &  80 & 12.747 & 0.0074  & 0.0024 & 0.0003 &   7.7 &     \\
GaSS 2-47 & 2022-12-27 & 2459940.684 &  30 &  45 & 12.754 & 0.0099  & 0.0045 & 0.0006 &   7.7 &     \\
GaSS 2-48 & 2022-03-02 & 2459640.582 &  20 & 120 & 12.091 & 0.0075  & 0.0024 & 0.0004 &   6.0 &     \\
GaSS 2-48 & 2023-01-04 & 2459948.679 &  20 &  35 & 12.029 & 0.0072  & 0.0026 & 0.0007 &   3.6 &     \\
GaSS 2-49 & 2022-02-25 & 2459635.675 &  90 &  15 & 13.520 & 0.0071  & 0.0031 & 0.0002 &  14.3 & a   \\
&&&\\
\hline 
\end{tabular}   
\tablefoot{The full table is only available electronically via CDS; a small
portion is shown here for guidance on its form and content.  The columns
list the (1) name, (2) date of observation, (3) median HJD of the observing
block, (4) exposure time of each $B$-band frame, (5) number of $B$-band
frames within the block, (6) median $B$-band magnitude of GaSS target during the
observing block, (7) $B$-band magnitude standard deviation of GaSS target (red star
in Fig.~\ref{fig:flickering}), (8) difference (dashed red vertical line in
Fig.~\ref{fig:flickering}) of the observed and expected magnitude standard
deviation (the latter being calculated based on GaSS magnitude and the fit
to check stars; i.e.,   dashed black line in Fig.~\ref{fig:flickering}), (9)
RMSE of the fit to check stars (the breadth of the gray band in
Fig.~\ref{fig:flickering}), (10) significance of our flickering detection
($\sigma_\mathrm{STD}$; see Eq.~(\ref{bor})), and (11) notes on the quality
of the data, where \textit{a} indicates a low number of check stars ($<$5)
and \textit{b} a poor distribution of check stars with brightnesses above and
below the SySt.}
\label{tab:flickering}
\end{table*}

In Paper~I, limited time allocation and unsupervised queued-robotic mode
restricted observations to a single $\sim$70min visit at a random epoch for
most of the targets.  Given the erratic nature of flickering in SySts
accreting at the lowest levels, and the variable photometric quality of
different nights (including seeing, guiding, and Moon phase), detecting
flickering in a single visit often depended on observing the right star at
the right time during a perfect night.  This likely accounts for the limited
number of flickering detections in Paper~I.

In contrast, all program stars in this paper were visited several times
between 2021 and 2024 (more than 9 times for eleven of them), increasing the
probability of collecting data during both optimal photometric conditions
and epochs of active flickering.  Furthermore, repeated observations allowed
us to refine the sampling times for subsequent visits and improve the
photometric comparison sequences and check star selection.  In a few cases,
this involved off-centering the target within the field to compensate for
the low stellar density encountered at high Galactic latitudes.  The only
drawback of the Asiago 67/92cm telescope (located at +46$^\circ$ latitude)
is the mechanical constraint preventing observations of targets south of
declination $-$23$^\circ$, a limitation affecting the four southernmost
program stars.

All targets were searched for flickering in the Landolt $B$ band, following
the same rationale as in Paper~I.  Although the entire optical train of the
Asiago 67/92 Schmidt telescope is UV-transparent (e.g., the corrector plate
is made of UBK7 glass), we did not attempt $U$-band observations.  This was
due to the prohibitive paucity of field stars (an order of magnitude less
than in $B$) and the significantly lower S/N, as the targets are typically
$\sim$2~mag fainter in $U$ than in $B$.  As demonstrated by the
\citet{2025A&A...701A.176M} radiative modeling of T~CrB during its 2005--2015
deep quiescence, $B$-band emission is dominated by the hot spot rather than
the accretion disk body or the red giant.  This makes the $B$ band ideal for
flickering observations of distant SySts accreting at low rates, as targeted
in this paper.

The flickering observations were calibrated individually from the
instantaneous local photometric system to the \citet{2009AJ....137.4186L}
standard system.  This was necessary to achieve and maintain millimag
precision over runs exceeding one hour.  A purely differential approach with
respect to field stars would not have reached the required precision, given
the instability of the atmosphere during the observations.  The
fractional-pixel stability of the telescope guiding during each run ensured
that the symbiotic star and all comparison and check stars remained on the
same pixels, minimizing flat-fielding errors and boosting internal accuracy. 
The observing strategy, data reduction, and selection of comparison and
check stars, along with the details of the aperture and PSF-fitting
photometry, are discussed in detail in Appendix~\ref{appendix:flickering}.

The results of our flickering analysis are summarized in
Table~\ref{tab:flickering}.  In Fig.~\ref{fig:flickering}, we highlight four
program stars (GaSS 2-43, 2-45, 2-46, and 2-49) exhibiting some of the most
prominent flickering detections, ranging from 0.0031 to 0.0057 mag (dashed
red lines).

To identify flickering, we calculated the standard deviation of the $B$
magnitude for each target (STD$_{\rm GaSS}$) within an observing block
($\sim$70 min) and compare it to the expected value of the standard
deviation (STD$_{\rm fit}$) for a non-flickering star of equivalent
magnitude.  The latter was derived from a second-order polynomial fit
(represented by the dashed black line in Fig.~\ref{fig:flickering}) to a
sample of check stars of similar red color to the program symbiotic star
(within $\Delta(B-V)$$\lesssim$0.2 mag) and located around it.  The
significance of the detection is quantified by $\sigma_\mathrm{STD}$,
which is defined as
\begin{equation}
    \sigma_\mathrm{STD} = \frac{\mathrm{STD_{GaSS}}-\mathrm{STD_{fit}}}{\mathrm{RMSE}}, 
\label{bor}
\end{equation}
where RMSE is the root mean squared error of the polynomial fit. 
$\sigma_\mathrm{STD}$ is equivalent to the ratio of the length of the red
dashed line to the width of the grey band in Fig.~\ref{fig:flickering}.  For
cases where $(\mathrm{STD_{GaSS}}-\mathrm{STD_{fit}})<0$, we uniformly set
$\sigma_\mathrm{STD}=-1$ to indicate the absence of detectable flickering. 
This metric effectively serves as a measure of the statistical significance
of our flickering detections.  The qualitative coding reported in
Table~\ref{tab:newsymb} corresponds to 2.5$\leq$$\sigma_\mathrm{STD}$$\leq$3.5,
3.5$\leq$$\sigma_\mathrm{STD}$$\leq$5.0, and $\sigma_\mathrm{STD}$$>$5.0 for
flags "y", "Y", and "Y!", respectively.

As illustrated in Sect.~\ref{RGflick} and the associated
Fig.~\ref{fig:checkstars}, the majority of the check stars are themselves
red giants.  Their stability over the $\sim$70 min flickering runs (black
points defining the gray bands of Fig.~\ref{fig:flickering}) proves that the
flickering observed in the program stars is not an intrinsic property of red
giants (either RGB or AGB), but is instead directly associated with
symbiotic activity.

\section{Near-UV excess in Gaia XP spectra}\label{XPspectra}

Gaia DR3 provides access to XP spectra, which are low-resolution, slitless
prism spectra covering the 3,300 to 10,500~\AA\ range in two arms---$B_P$
and $R_P$---with minimal overlap at H$\alpha$.  The XP spectra are provided
in a flux-calibrated format as the mean over all individual spectra
collected during the first 34 months of satellite operations (2014--2017). 
Technical details are summarized in Appendix~\ref{appendix:XPspectra}.

\begin{figure*}
   \centering
   \includegraphics[angle=270,width=17cm]{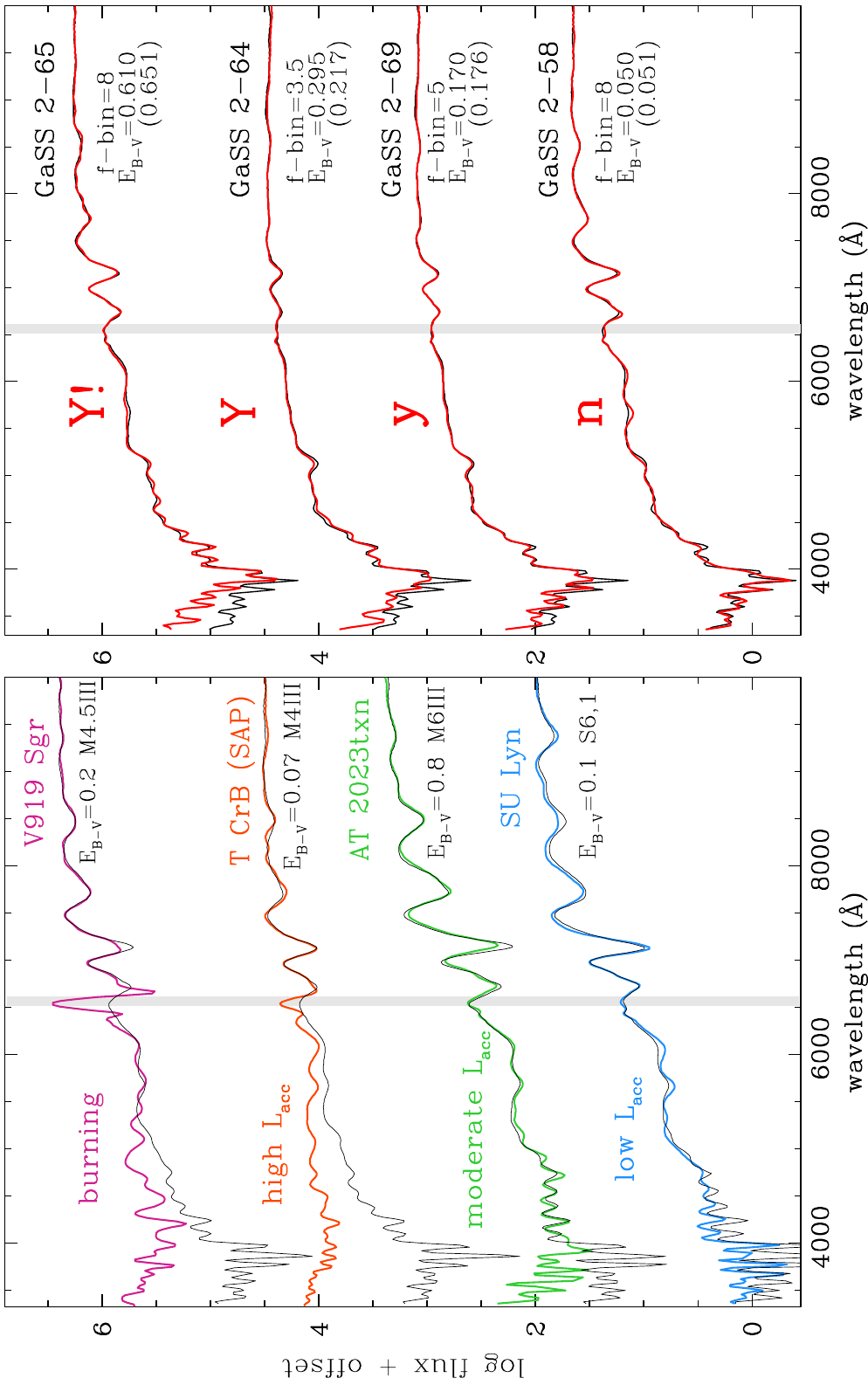}
      \caption{
        Examples of reddening-corrected Gaia XP spectra for well-known
        symbiotic stars (left panel) and new symbiotic stars discovered in
        this work (right panel).  The best-fit spectral templates (see
        Appendix~\ref{appendix:XPspectra}) and the applied reddening
        corrections \citep[following][]{1999PASP..111...63F} are indicated
        for each object \citep[in parentheses, we show the reddening values from the
        maps of][]{1998ApJ...500..525S}.  The vertical gray band marks the
        position of H$\alpha$.  The symbiotic stars in the left panel
        illustrate typical XP spectra for H-burning conditions as well as
        for high, moderate, and low accretion luminosities.  Those in the
        right panel illustrate the varying degrees of near-UV excess
        ($\lambda \leq 4000$~\AA) of the program's new SySts and the
        corresponding classification (n, y, Y, and Y!) adopted in
        Table~\ref{tab:newsymb}.}
         \label{fig:XPexamples}
   \end{figure*}

As an example of their diagnostic potential, in the left panel of
Fig.~\ref{fig:XPexamples} we compare the XP spectra of four symbiotic stars
with those representative of their red giant components (see
Fig.~\ref{fig:XPsymbio}).  The four selected symbiotic stars exemplify
those well: ($i$) in stable H-burning (V919 Sgr); ($ii$) during high
accretion states (e.g., T CrB in 2015--2017); ($iii$) accreting at a moderate
rate (AT 2023txn); and ($iv$) accreting at low or very low levels (SU Lyn). 
Note that the apparent noise shortward of 4000~\AA\ is not intrinsic to the
source but is a manifestation of the Gaia DR3 choice to represent spectra
using Hermite functions (see  Appendix~\ref{appendix:XPspectra} for a
discussion).

As clear from Fig.~\ref{fig:XPexamples}, a large near-UV excess
characterizes both burning and high-accretion types; in the latter, the
excess emission extends to redder wavelengths, while a much stronger
H$\alpha$ line is typical of the former.  No significant H$\alpha$ emission
is detectable in XP spectra for moderate and low accretion types, which
differ primarily by the amount of near-UV excess emission shortward of
4000~\AA.  Despite the low resolving power, Gaia XP spectra are sufficient
to resolve some molecular bands and provide a spectral classification of the
red giant.  In addition, it is possible to detect chemical anomalies such as
the S6,1 spectrum for the red giant in SU Lyn \citep[note the ZrO 9330~\AA\
band, which is typical of S-type stars;
see][]{1976ims..book.....P,2009MNRAS.400.1413W}, a star traditionally
considered a normal M6\,III \citep{2016MNRAS.461L...1M}.  Simultaneous with
the spectral typing of the red giant, the reddening ($E_{B-V}$) can also be
well constrained thanks to the wide wavelength coverage of the XP spectra.

For the low accretion rates of interest in this series of papers, H$\alpha$
emission is too weak to provide a detectable signature in XP spectra. 
Consequently, the most viable indicator of an interacting binary nature is
the presence of excess flux at near-UV wavelengths.  To detect such an
excess, it is necessary to compare the reddening-corrected spectra of the
program stars with suitable templates.  To build templates for all $f$-bins,
we have median-combined the XP spectra of 32,182 M giants ("32k sample"
hereafter) that we identified among the one~million stars observed by
Galah so far.  We limited the selection to the 32k stars with viable XP
spectra and with a reddening of $E_{B-V}$$<$0.1~mag as derived from
\citet{1998ApJ...500..525S} maps (see details in
Appendix~\ref{appendix:XPspectra}).

The spectra of the newly discovered SySts were iteratively fitted to such
$f$-bin templates while simultaneously adjusting the reddening correction
following the extinction law of \citet{1999PASP..111...63F}.  The reddening
values returned by the fitting procedure align well with those derived from
the \citet{1998ApJ...500..525S} maps; this consistency enhances confidence
in the methodology and suggests that any circumstellar reddening, if
present, is modest.  The fitting was restricted to $\lambda$$\geq$4200~\AA\
to avoid circularity with the search for flux excess at
$\lambda$$\leq$4000~\AA.  This procedure yielded excellent matches, examples
of which are provided in the right panel of Fig.~\ref{fig:XPexamples}.

The presence of a UV excess was then evaluated by comparing the XP spectra
of the matching template and the program star for $\lambda$$\leq$4000~\AA,
grading it into four classes: "y" pertains to stars for which data points
shortward of 4000~\AA\ show a mean excess between 3 and 5$\sigma$ with
respect to the template spread; class "Y" for 5 to 7$\sigma$; and "Y!" for
excesses larger than 7$\sigma$.  Examples for each class, including the case
of no excess ("n"), are plotted in the right panel of
Fig.~\ref{fig:XPexamples}.  The final results, including the matching
$f$-bin, reddening, and near-UV excess classification, are summarized in
Table~\ref{tab:newsymb}.

   \begin{figure}
   \centering
   \includegraphics[width=8.8cm]{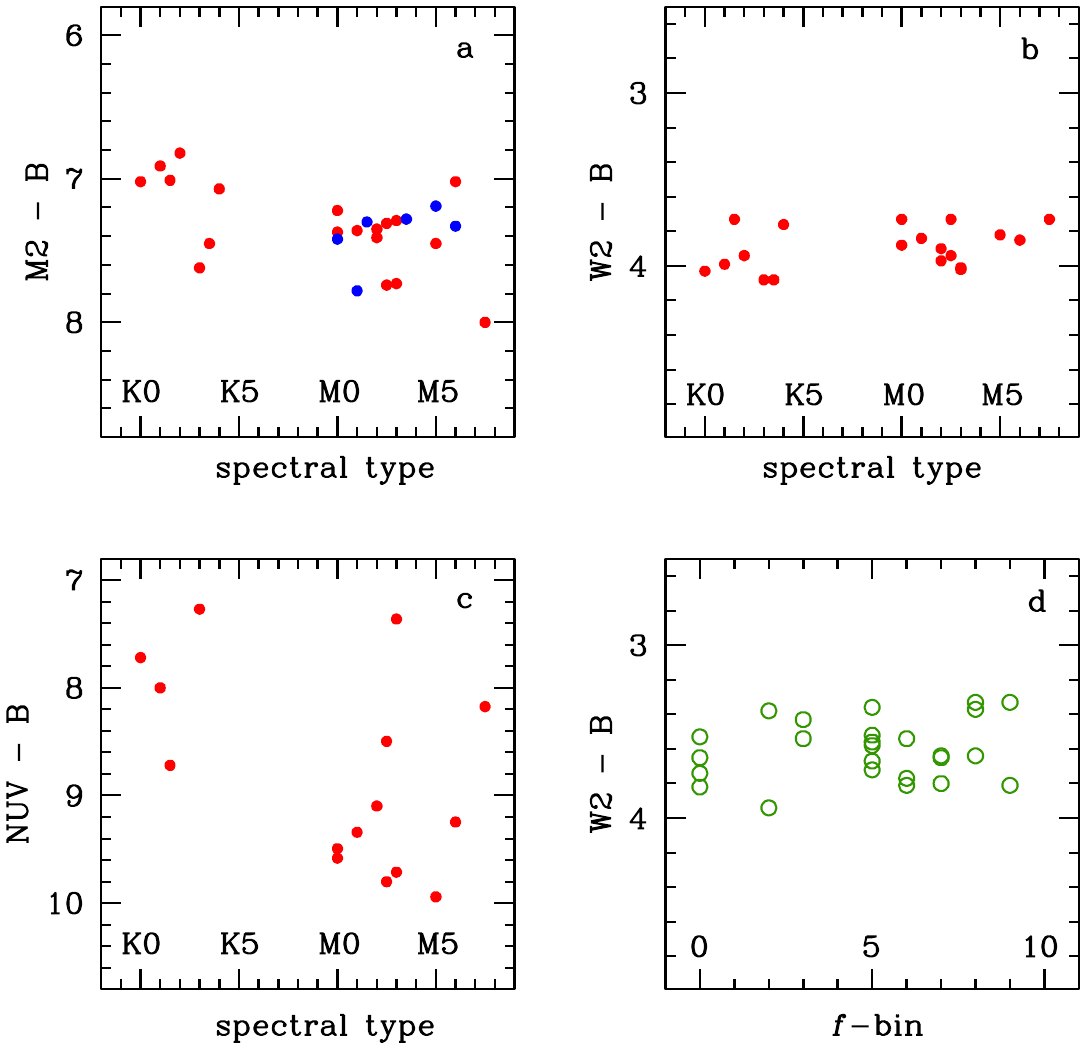}
      \caption{Photospheric emission of normal field giants in the {\it
      Swift} M2 and W2 ultraviolet bands (panels $a$ and $b$), and GALEX NUV
      band (panel $c$), expressed as magnitude difference with the optical
      $B$ band (data from Table~\ref{tab:appendixSwift}; for details and the
      meaning of blue points, see Sect.~\ref{Swift}).  For the new SySts,
      panel $d$ plots the W2$-B$ color derived from data in
      Tables~\ref{tab:newsymb} and \ref{tab:Swift}.}
         \label{fig:M2W2B}
   \end{figure}

\section{Swift X-ray and UV observations}\label{Swift}

All program stars were observed in X-rays and in the UV with
the {\it Swift} satellite \citep{2004ApJ...611.1005G}.  Between March, 2022
and July, 2024, we observed 37 of the 44 program stars through two Guest
Observer programs (IDs 1821023 and 1922036; PI: N.  Masetti) during {\it
Swift} cycles 18 and 19.  Moreover, between February and June, 2024, five
target-of-opportunity (ToO) observations of sources GaSS 2-53, GaSS 2-55,
GaSS 2-64, GaSS 2-68, and GaSS 2-71 were acquired with the same satellite. 
For three additional objects (GaSS 2-40, GaSS 2-48, and GaSS 2-57), a number
of serendipitous observations were retrieved from the ASI-{\it Swift}
archive\footnote{{\tt http://www.ssdc.asi.it}}.  Details of the data
reduction and analysis are provided in Appendix~\ref{appendix:Swift}, and the
results are listed in Table~\ref{tab:Swift}.

Eight targets were also detected in the UV by the GALEX satellite
\citep{2005ApJ...619L...1M} during its all-sky survey performed between
April, 2003 and June, 2013. Their magnitudes in the near-UV (NUV) filter
\citep{2017ApJS..230...24B} are reported in Table~\ref{tab:GALEX}. None of
the program stars were detected by GALEX in the far-UV (FUV)  band.

\subsection{The nature of the accreting source}

Given the low accretion luminosity (as inferred from the H$\alpha$ emission;
see Sect.~\ref{luminosity}) and the mean distance of 3.2~kpc (see
Sect.~\ref{distance}), we did not expect to detect X-ray emission from our
targets, except in two cases: (1) a given program star was undergoing a
phase of greatly enhanced accretion at the time of the {\it Swift}
observations, or (2) the accreting degenerate star is a neutron star instead
of a white dwarf.  The potential well of a neutron star is so deep that even
very modest accretion rates (lower than those characterizing the program
stars) would have resulted in a detectable hard X-ray source, as illustrated
by the observations in \citet{2007A&A...470..331M} and the modeling in
\citet[][see their Figure 7]{2022A&A...661A.124M}.

None of the 44 new SySts were positively detected in X-rays.  We conclude
that none of them were undergoing a surge in accretion at the time of
observations, and none harbor a neutron star; this confirms the rarity of
neutron stars compared to white dwarfs as the degenerate accretors in SySts.

\begin{table*}
\caption{Results of the analysis of X--ray and UV data collected with the
{\it Swift} satellite on all of the 44 program stars.  Errors are at 1-$\sigma$
confidence level; upper limits refer to the 3$\sigma$ confidence level.}
\label{tab:Swift}
\begin{tabular}{lcc|rll|rrcc|l}
\hline\hline
       &              
       &           
       &
\multicolumn{3}{c|}{XRT}&
\multicolumn{4}{c|}{UVOT~~}&
       \\
Object &    
Date   &   
start  & 
\multicolumn{1}{c}{expt} & 
\multicolumn{1}{c}{count}& 
\multicolumn{1}{c|}{flux$^b$} & 
\multicolumn{1}{c}{N} & 
\multicolumn{1}{c}{expt}  & 
UVM2 & 
UVW2 & 
Remarks \\
       &
       & 
(UT)   &
(sec)  & 
\multicolumn{1}{c}{rate$^a$}  & 
       &
\multicolumn{1}{c}{} & 
\multicolumn{1}{c}{(sec)} &  
(mag)  &  
(mag)  &       
       \\
\noalign{\smallskip}
\hline
GaSS 2-34 & 23 May 2023  &   01:56   &   950   &  $<$1.5  &  $<$1.3  &  3  &  935  &                &  16.36$\pm$0.04  & $c$ \\
GaSS 2-34 & 19 May 2024  &   15:51   &  2455   &          &          &  5  & 2433  & 19.55$\pm$0.13 &                  & $d$ \\
GaSS 2-34 & 25 May 2024  &   02:42   &   993   &          &          &  1  &  979  & 19.53$\pm$0.19 &                  & \\
GaSS 2-35 & 03 Apr 2022  &   08:45   &   584   &  $<$6.0  &  $<$5.3  &  1  &  576  &                &  17.97$\pm$0.09  & \\
GaSS 2-35 & 13 Jul 2022  &   15:20   &   522   &          &          &  1  &  514  &                &  18.05$\pm$0.10  & \\
GaSS 2-36 & 01 Apr 2022  &   12:17   &   489   &  $<$1.1  &  $<$1.0  &  2  &  454  &                &  18.44$\pm$0.12  & $c$ \\
GaSS 2-36 & 03 Apr 2022  &   04:16   &  4739   &          &          & 10  & 4614  &                &  18.45$\pm$0.05  & \\
GaSS 2-36 & 24 Jul 2023  &   19:54   &   657   &          &          &  1  & 646   &                &  18.62$\pm$0.12  & \\
GaSS 2-37 & 31 Mar 2022  &   01:26   &  1918   &  $<$3.4  &  $<$3.0  &  3  & 1886  &                &  18.73$\pm$0.08  & $c$ \\
GaSS 2-38 & 11 Apr 2023  &   19:21   &   918   &  $<$4.5  &  $<$4.0  &  1  &  912  &                &  17.44$\pm$0.07  & $e$ \\
GaSS 2-38 & 12 Apr 2023  &   14:33   &   542   &          &          &  2  &  532  &                &  17.19$\pm$0.07  & \\
GaSS 2-39 & 13 Apr 2023  &   15:56   &  2026   &  $<$3.3  &  $<$2.9  &  2  & 1992  &                &  18.15$\pm$0.07  & $c$ \\
\noalign{\smallskip}
\hline
\hline
\noalign{\smallskip}
\end{tabular}
\tablefoot{The full table is only available electronically via CDS; a small
portion is shown here for guidance on its form and content. 
N = number of exposures;
$^a$ = 0.3$-$10 keV count rate ($\times$10$^{-3}$\,s$^{-1}$);
$^b$ = 0.3$-$10 keV flux ($\times$10$^{-13}$ erg cm$^{-2}$ s$^{-1}$);
$c$ = no variability across UVW2 exposures; 
$d$ = tentative variability across UVM2 frames; 
$e$ = 2.5$\sigma$ variability between UVW2 observations on adjacent days.
}
\end{table*}

\begin{table}
\caption[]{Detection by GALEX in the NUV band of eight program stars.}
\label{tab:GALEX}
\begin{center}
\begin{tabular}{lc|cc}
\hline\hline
\noalign{\smallskip}
\multicolumn{1}{c}{Object} & NUV mag. & Object & NUV mag.\\
\noalign{\smallskip}
\hline
\noalign{\smallskip}
GaSS 2-34  & 20.92$\pm$0.23 & GaSS 2-47  & 21.23$\pm$0.28 \\
GaSS 2-39$^\dagger$  & 22.39$\pm$0.50 & GaSS 2-48  & 20.64$\pm$0.13 \\
GaSS 2-42  & 21.82$\pm$0.31 & GaSS 2-74  & 22.46$\pm$0.39 \\
GaSS 2-44  & 22.45$\pm$0.34 & GaSS 2-77  & 22.15$\pm$0.34 \\
\hline
\end{tabular} 
\tablefoot{The magnitudes are in the native AB system.  $\dagger$: Possible
spurious detection due to its proximity with the edge of the instrument's
field of view.}
\end{center}
\end{table}  

\subsection{Ultraviolet signature of accretion}\label{UVOT}

To ascertain the presence of excess emission in the {\it Swift} UVOT
observations of the 44 new SySts in the M2 and W2 bands, it is first
necessary to derive the level of emission expected from the photosphere of
the cool giant alone and then compare it to the recorded data.  To this end,
we searched the {\it Swift} archive for serendipitous observations of field,
high Galactic latitude, and non-variable red giants with accurate spectral
classifications from \citet{1999mctd.book.....H}.  Such giants are
sufficiently faint to avoid overloading the {\it Swift} detectors.  We found
16 suitable red giants in the archive, which we supplemented with ToO
observations of four additional objects (all of M spectral type).  The
results are summarized in Table~\ref{tab:M2W2B}.

The level of photospheric emission in the UVOT M2 and W2 bands from these
red giants can be conveniently expressed as a magnitude difference with
respect to the optical $B$-band, which circumvents uncertainties related to
distance.  The (W2$-$$B$) and (M2$-$$B$) color indices computed from
Table~\ref{tab:M2W2B} are plotted as a function of the spectral type in
Fig.~\ref{fig:M2W2B}, where we added the results for six further M-type
giants obtained from their ultraviolet IUE spectra (blue points in
Fig.~\ref{fig:M2W2B}).  For these six giants, we combined their IUE spectra
with 3200--8000~\AA\ spectra obtained with the Asiago 1.22m telescope,
scaled to match the IUE flux over the 3200--3350~\AA\ overlapping region. 
This procedure eliminates the effects of stellar variability between
non-simultaneous observing epochs.  The (M2$-$$B$) color index was then
computed by direct integration of the passband profiles taken from
\citet{2008MNRAS.383..627P} and \citet{1992AJ....104..340L}, with flux
zero-points set on Vega.

\subsubsection{Swift M2 band}

Since the M2 and $B$ bands lie well within the Wien tail of the energy
distribution for red giants, it is expected that the (M2--$B$) index is
similar for K- and M-type giants, as confirmed by panel~$a$ of
Fig.~\ref{fig:M2W2B}.  For M-type giants, the mean value is
$\langle$(M2--$B$)$\rangle$=7.44 with a dispersion $\sigma$=0.23; therefore,
the 3$\sigma$ limit for detecting excess emission in our program stars is
(M2--$B$)$\leq$6.70~mag.  The results in Table~\ref{tab:Swift} show that
only three program stars were detected in the M2 band out of the 24 observed
(a 13\% success rate).  GaSS~2-52 and GaSS~2-54, with (M2--$B$) values of
5.65 and 5.02 respectively, offer a robust detection of excess emission
originating from the accretion process.  For GaSS 2-34 we find
(M2--$B$)=6.65, which is just over the 3$\sigma$ limit, meaning a highly
probable detection of an ultraviolet excess.

\subsubsection{GALEX NUV band}

The GALEX NUV band is centered at a similar wavelength to {\it Swift} M2. 
Most of the field red giants listed in Table~\ref{tab:M2W2B} were detected
by GALEX in the NUV band, allowing us to explore the (NUV$-$$B$) index
similarly to the (M2$-$$B$) considered above.  The (NUV$-$$B$) values
plotted in panel $c$ of Fig.~\ref{fig:M2W2B} are, however, scattered all
over the place, which suggests that GALEX NUV data are not very useful in
defining a meaningful reference level for the photospheric emission of the
red giants or determining if any of the program stars in
Table~\ref{tab:GALEX} show excess emission originating from the accretion
process.

\subsubsection{Swift W2 band}

The low detection rate in the M2 band was expected given the low accretion
luminosity and large distances.  This prompted us to explore the W2 band
despite the known red-leak affecting its transmission profile.  The
effective central wavelength of the W2 band is 2030~\AA, compared to
2220~\AA\ for M2, which means a reduction of 2$\times$ in the photospheric
emission of the red giant.  In addition, the M2 filter is affected by a
somewhat lower peak transmission than W2, and within its passband lies the
strong 2175~\AA\ hump characterizing the interstellar extinction law.

We hypothesized that hot accretion regions, probed at W2, could provide a
signal detectable above the red leak.  At the time of the observations, in
the literature there was no quantitative observational assessment of the
red leak specifically for red giants.  Simulations provided limited guidance
because the tabulated transmission in \citet{2008MNRAS.383..627P} ends well
before the extended red-tail sensitivity of the detector.

As with the M2 and NUV bands above, the data in Table~\ref{tab:M2W2B} were
used (panel $b$ of Fig.~\ref{fig:M2W2B}) to define the (W2--$B$) value for
pure photospheric emission from the red giant.  The distribution appears
quite tight, with a mean $\langle$(W2--$B$)$_{std}$$\rangle$=3.87 and
$\sigma_{std}$=0.10 for the stars of M spectral type.

In the W2 band we observed 26 of the 44 new SySts and detected all of them.
The results are summarized in Table~\ref{tab:Swift} and plotted in
panel $d$ of Fig.~\ref{fig:M2W2B}.  They show
$\langle$(W2--$B$)$_{SySts}$$\rangle$=3.64 with $\sigma_{SySts}$=0.17.

While individual assessments are difficult, the mean value for SySts is
bluer than the reference M-type giants by 0.23~mag (corresponding to
2.3$\sigma_{std}$), a statistically significant difference, suggesting that
an ultraviolet excess produced by the accretion processes can be detected
(although faintly) over the W2 red leak for the population of low-accretion
SySts explored in this paper.  For the sake of completeness, we note that
for the M-type stars in Table~\ref{tab:M2W2B}, the W1 filter (of effective
central wavelength 2634 \AA\ and also strongly affected by red-leak)
provides $\langle$(W1--$B$)$_{std}$$\rangle$=2.93 with $\sigma_{std}$=0.05.

   \begin{figure*}
   \includegraphics[angle=270,width=18cm]{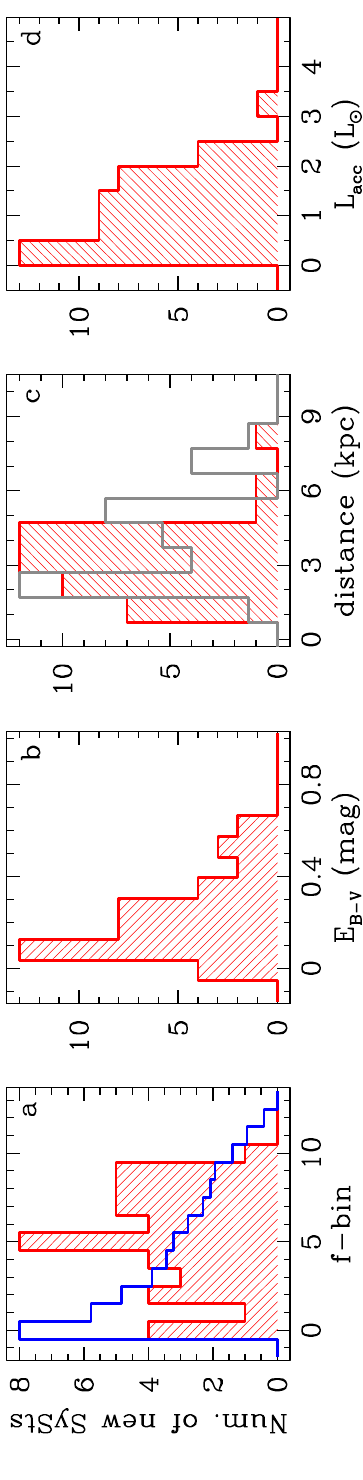}
        \caption{Distribution of the 44 new symbiotic stars in terms of (a)
        $f$-bin classification of the red giant (the blue histogram
        represents the distribution of the full sample of 32k M-giants
        observed by Galah); (b) interstellar reddening
        \citep[from][]{1998ApJ...500..525S}; (c) distance (derived from
        absolute $R_P$ and $K_s$ magnitudes of the red giant; that from Gaia
        DR3 parallax when the error is $<$20\% is shown in gray); and (d)
        mean accretion luminosity (computed from the mean H$\alpha$ emission
        flux; see Sect.~\ref{discussion} for details).  Data from
        Table~\ref{tab:summary}.}
        \label{fig:histograms}
   \end{figure*}

\section{Discussion}\label{discussion}

\subsection{Distribution of spectral types}

It is an established fact that the spectral types of red giants in
well-known SySts are skewed toward cooler subtypes
compared to the distribution of field giants
\citep[e.g.,][]{1986syst.book.....K}, with a peak around M5--M6~III
\citep{1999A&AS..137..473M}.  This is typically attributed to the higher
mass-loss rates of cooler red giants, which favor increased accretion onto
the white-dwarf companions.

The $f$-bin distribution of the red giant components for the 44 new SySts
discovered in this work is shown in the panel~$a$ of
Fig.~\ref{fig:histograms}, where it is compared with the distribution of the
full sample of 32k M-giants observed by Galah.  The distribution of these
new SySts is clearly shifted toward cooler f-bins, consistent with the trend
observed in the well-known SySt population.  This supports the view
advocated by \citet{2019arXiv190901389M} that the subdivision of symbiotic
stars into burning-types and the various accreting-only states (high,
moderate, and low) does not necessarily reflect fundamental differences in
the underlying binary population.  Instead, these categories likely
represent different evolutionary states that each symbiotic star cycles
through during its lifetime.

\subsection{Great variability of the accretion levels}

\begin{figure}
   \includegraphics[width=\columnwidth]{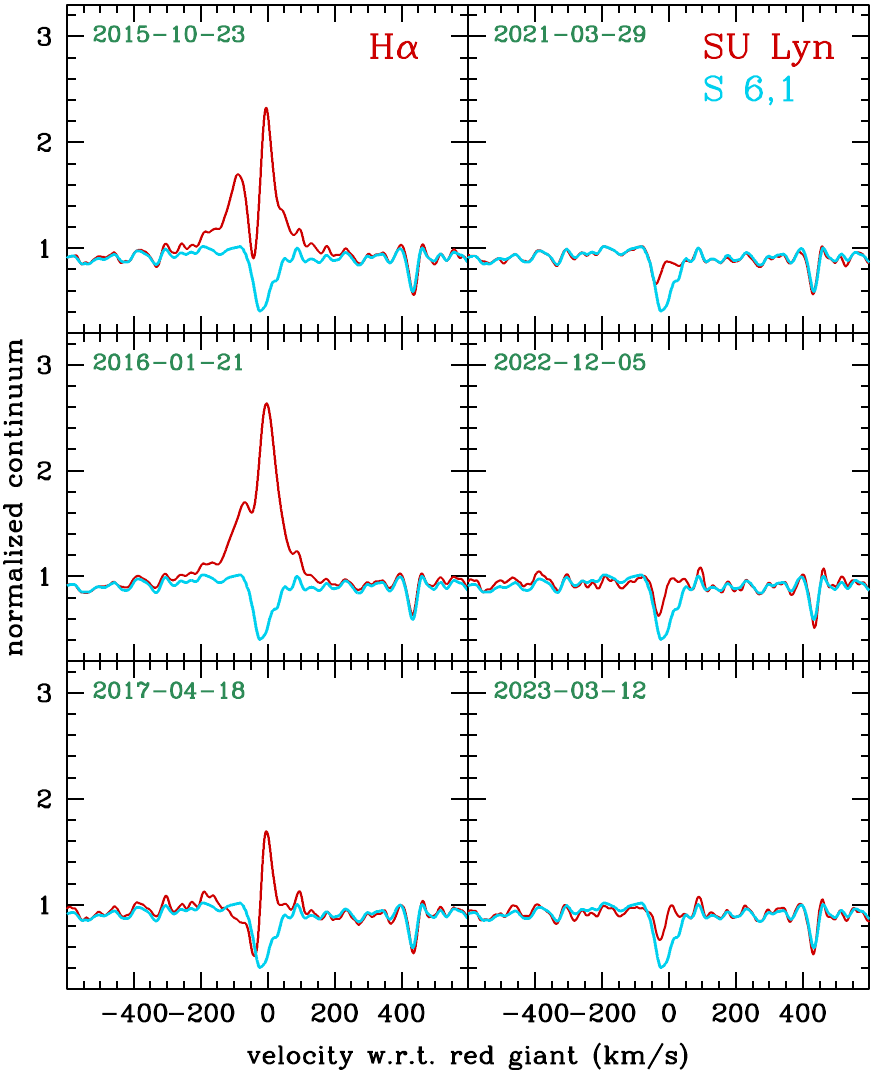}
        \caption{Profile and flux variability of the H$\alpha$ emission line
        in SU~Lyn, the prototype of LowAcc symbiotic stars. 
        High-resolution spectra of SU Lyn are compared to the spectrum of
        HD~284438, an S-type giant of S6,1 spectral type
        \citep[following][]{1954ApJ...120..484K}.  All spectra were obtained
        with the Asiago 1.82m telescope equipped with the Echelle
        spectrograph.}
        \label{fig:SULyn}
   \end{figure}

When SU Lyn---the prototype of LowAcc symbiotic stars---was discovered in
2015–2016 by \citet{2016MNRAS.461L...1M} as the counterpart of an
unidentified hard X-ray source, its identification was made possible because
the object was then experiencing suitably high levels of UV excess (see 
Fig.~\ref{fig:XPexamples}) and H$\alpha$ emission (see 
Fig.~\ref{fig:SULyn}).  We have since continuously monitored the near-UV
excess and H$\alpha$ emission of SU Lyn (Munari et al., in prep.), finding
that both have remained significantly below the 2015--2016 levels, as
illustrated for several epochs in Fig.~\ref{fig:SULyn}.  At the accretion
levels displayed by SU Lyn from 2019--2023 (roughly 1/20th of the 2015--2016
values), the object would likely have escaped identification, and the
presence of a population of LowAcc SySts might have remained unsuspected.

All 44 new symbiotic stars discovered and characterized in this paper
exhibit a similarly large variability in their accretion levels, closely
resembling the behavior of the prototype SU Lyn.  As illustrated by GaSS
2-35 in Fig.~\ref{fig:Ha_evolution}, the H$\alpha$ emission can vary
dramatically over time, by a factor of $\sim$20 between 2022 Nov 4 and 2024
Aug 21. While H$\alpha$ emission is "off" most of the time for GaSS 2-40 and
GaSS 2-67, mirroring the behavior of SU Lyn, it is typically "on" in GaSS
2-35 and GaSS 2-43, with the rest of the program stars distributed in
between the two.

\begin{figure}
   \includegraphics[width=\columnwidth]{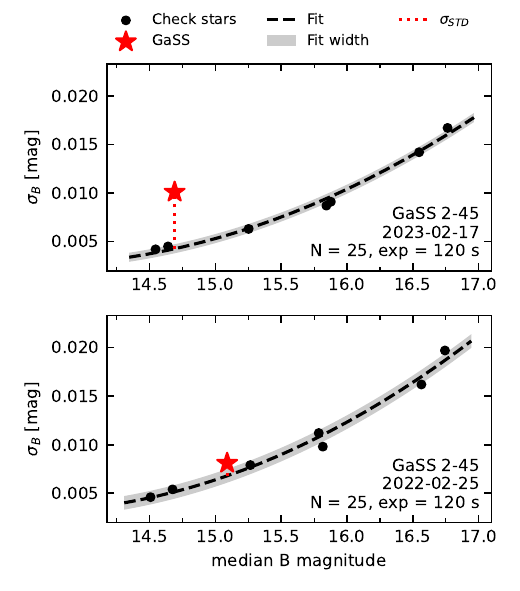}
        \caption{Variable presence of flickering in GaSS 2-45 for very
        similar observations taken one year apart (see
        Fig.~\ref{fig:flickering} for explanation of symbols).}
        \label{fig:flickering_GaSS_2_45}
   \end{figure}

The presence and amplitude of flickering are also highly variable and,
similarly to H$\alpha$ emission, support a large variability in the
accretion levels experienced by the newly discovered SySts.  For example,
Fig.~\ref{fig:flickering_GaSS_2_45} compares the search for flickering in
GaSS 2-45 across two distinct epochs: 2022 Feb 25 and 2023 Feb 17.  The
observations were conducted identically on both dates (120 sec integrations
for 25 data points in the $B$ band, covering a total duration of 70 min). 
The night quality -- considering seeing, guiding, and atmospheric stability
-- was similar, as evidenced by the equivalent distribution of the check
stars defining the distribution of field red stars (gray band in
Fig.~\ref{fig:flickering_GaSS_2_45}).  Nevertheless, GaSS 2-45 displayed
significant flickering on 2023 Feb 17 (amplitude = 5.7 mmag), which was
instead absent on 2022 Feb 25 (amplitude < 0.9 mmag).

When Swift/UVOT observations were executed as a series of separate
exposures, the results for both the M2 and W2 filters suggest
variability in excess of that attributable to the red giant's photospheric
emission (see the Remarks column in Table~\ref{tab:Swift}).  This likely
reflects the stochastic nature of accretion onto the degenerate companion.

In summary, LowAcc SySts are elusive, and recognizing them requires
accurate, protracted, and multi-technique observations, as exemplified by
GaSS 2-76.  In addition to deriving a well-defined binary orbit, our
2021--2025 observations of GaSS 2-76 have revealed large-amplitude
flickering and intense and highly variable H$\alpha$ emission; however, the
Gaia DR3 XP observations from a decade earlier (2014--2017) show at most a
feeble near-UV excess, through which the object would not have been
distinguished from normal RGs.

\subsection{Reddening and distances}\label{distance}

The amount of interstellar $E_{B-V}$ reddening affecting the new symbiotic
stars has been derived from the Galactic dust maps of
\citet{1998ApJ...500..525S}.  It is listed in Table~\ref{tab:summary} and
its distribution is shown in the panel~$b$ of Fig.~\ref{fig:histograms}. 
The reddening is generally small---as expected given the Galah footprint,
which avoids the Galactic plane---and is well-constrained.  This is
evidenced by the close agreement between the values estimated from the
\citet{1998ApJ...500..525S} maps and those obtained by fitting the Gaia XP
spectra (see the $E_{B-V}$ values compared in the right panel of
Fig.~\ref{fig:XPexamples}).  This consistency suggests that any
circumstellar contribution is generally negligible, as is typical of SySts with
a non-pulsating RG.

Distances to all program stars were derived by comparing the observed $R_P$
and $K_s$ magnitudes to the absolute magnitudes listed in
Table~\ref{tab:absmag}.  We applied reddening corrections following the
\citet{1999PASP..111...63F} extinction law, assuming the standard case of
$R_V$=3.1.  The resulting distance distribution is shown in the panel~$c$ of
Fig.~\ref{fig:histograms}, where it is compared with distances derived from
the direct inversion of Gaia DR3 parallaxes (for cases where the formal
parallax error is $<$20\%).  The two distributions are similar, with mean
values of 3.2 and 3.9~kpc for the photometric and parallax-based distances,
respectively.

\subsection{Accretion luminosities}\label{luminosity}

It is common practice to relate the bolometric accretion luminosity to the
luminosity radiated in a single emission line using power-law relations of
the following form: $\log(L^{\rm bol}_{\rm acc}) = \alpha_{\rm line} +
\beta_{\rm line} \log(L_{\rm line})$. Most of the available relations focus
on the H$\alpha$ emission line and were calibrated using accretion disks of
protostars observed over a wide range of stellar masses
\citep[e.g.,][]{1998AJ....116.2965M, 2008ApJ...681..594H, 2011A&A...535A..99M,
2017A&A...600A..20A, 2025A&A...704A..42F}.  An average of these relations
(in solar units) is given by
\begin{equation}
\log(L^{\rm bol}_{\rm acc}) = 2.23(\pm 0.25) + 1.1(\pm 0.12) \log(L_{\rm H\alpha}) 
\label{equat}
,\end{equation}
which is very close to the relation by \citet{2011A&A...535A..99M} adopted
in Paper I.

Very few bolometric luminosities are available in the literature for
accreting-only symbiotic stars, and these generally lack a simultaneous
determination of the H$\alpha$ luminosity.  Consequently, a direct
recalibration of Eq.~(\ref{equat}) specifically for SySts is currently not
feasible.  Instead, a more readily accessible quantity is the accretion
luminosity radiated longward of 2000~\AA\ ($L^{\rm \lambda>2000~\AA}_{\rm
acc}$), which is obtained through coordinated $UBVRI$ photometry, optical
spectrophotometry (that includes H$\alpha$), and ultraviolet observations in
the W2, M2, and W1 bands with the {\it Swift} satellite.  By combining our
data with literature values for a dozen measurements of accreting-only
SySts, we derived the following preliminary relation (which we plan to
investigate thoroughly in a future study):
\begin{equation}
\log(L^{\rm \lambda>2000~\AA}_{\rm acc}) = 1.5(\pm 0.2) + 1.1(\pm 0.1) \log(L_{\rm H\alpha}) 
.\end{equation}
This relation confirms the slope of Eq.~(\ref{equat}) and is consistent with
its intercept, considering that the energy radiated below 2000~\AA\ does not
contribute to $L^{\rm \lambda>2000~\AA}_{\rm acc}$.  This close agreement
suggests that, in continuity with Paper I, it is reasonable to use
Eq.~(\ref{equat}) to estimate the bolometric luminosity of the program stars
in this work, even though Eq.~(\ref{equat}) was formally calibrated on
protostellar accretion disks.

The epoch H$\alpha$ fluxes listed in Table~\ref{tab:Ha_evolution} were used
to compute the mean and peak values reported in Table~\ref{tab:summary}. 
These values were then converted, using Eq.~(\ref{equat}), into the mean and peak
accretion luminosities listed in Table~\ref{tab:summary}.  Finally, the
distribution of the 44 new SySts as a function of their mean accretion
luminosity is shown in panel $d$ of Fig.~\ref{fig:histograms}.

The mean value of the accretion luminosity for the new SySts is confined to
0.25--2.4~L$_\odot$ (5\%--95\% range), a value similar to the amount
radiated by T~CrB during deep quiescence \citep[see][]{2025A&A...701A.176M},
and which also compares well with the range displayed by SU Lyn, the
prototype of LowAcc SySts.  In fact, the accretion luminosity derived by
\citet{2016MNRAS.461L...1M} for SU Lyn from 2015--2016 was
$\sim$10~L$_\odot$, at a time when the H$\alpha$ emission was higher than ever
observed in any of the new SySts of this paper (see  Fig.~\ref{fig:SULyn}),
while a minimum value of $\sim$0.5~L$_\odot$ was probably reached by SU Lyn
from 2019--2023, when the intensity of H$\alpha$ dropped to $\approx$1/20 of
that in 2015--2016 (see  Fig.~\ref{fig:SULyn}). 

The mean accretion luminosity for the newly identified SySts is lower than
that of the objects in Paper~I, which were characterized by either stronger
H$\alpha$ emission or X-ray detections.
In Table~\ref{tab:summary} we also quote the flux upper limits that have
been computed from the X-ray count rates in the 0.3--10 keV band reported in
Table~\ref{tab:Swift} using the WebPIMMS\footnote{{\tt
https://heasarc.gsfc.nasa.gov/cgi-bin/ Tools/w3pimms/w3pimms.pl}} online
tool and assuming the X-ray spectrum of SU Lyn as in
\citet{2016MNRAS.461L...1M}, i.e.,  an absorbed thermal plasma with
a temperature of $kT$ = 17 keV and hydrogen column density of $N_{\rm H}$ =
2.9$\times$10$^{22}$ atoms cm$^{-2}$.  The corresponding luminosity limits
were calculated according to the distances reported in
Table~\ref{tab:summary}.  These limits (expressed in the above table in
solar luminosity units) refer to absorbed values for the X-ray fluxes; the
unabsorbed ones are $\sim$50\% larger.  These upper limits to L$_{\rm X}$
are fully consistent with accretion luminosity derived from the flux in
H$\alpha$.

   \begin{figure}[t!]
   \includegraphics[width=\columnwidth]{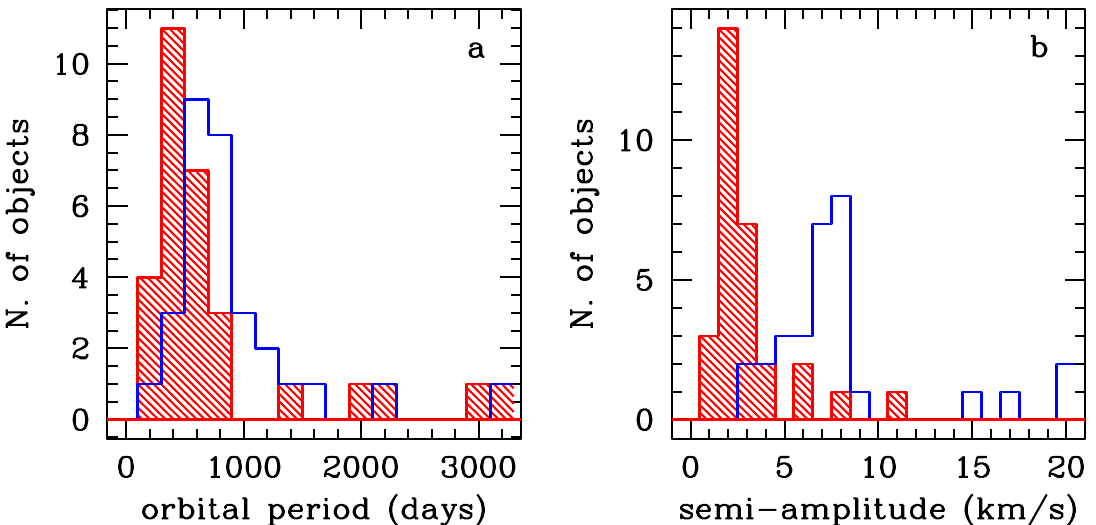}
        \caption{Distribution in orbital period and in semi-amplitude of the
        spectroscopic orbits of the new SySts listed in Table~\ref{tab:orbits}.
        For comparison the distributions for the Fekel\,\&\,Hinkle sample 
        defined in Sect.~\ref{FekelHinkle} are shown in blue.} 
        \label{fig:PorbKamp}
   \end{figure}

\subsection{Comparing the orbits to well-known symbiotic
stars}\label{FekelHinkle}

The largest and most homogeneous set of spectroscopic orbits for
well-known SySts is undoubtedly that collected at infrared wavelengths over
twenty years by \citet{2000AJ....119.1375F, 2000AJ....120.3255F,
2001AJ....121.2219F, 2007AJ....133...17F, 2008AJ....136..146F,
2010AJ....139.1315F, 2015AJ....150...48F, 2017AJ....153...35F} and
\citet{2006ApJ...641..479H, 2009ApJ...692.1360H}, hereafter referred to as
the Fekel\,\&\,Hinkle sample. This sample comprises 30 objects selected
from the brightest SySts near the Sun, generally at low Galactic latitudes
and therefore lying close to the Galactic plane. In contrast, our sample
of 30 spectroscopic orbits in Table~\ref{tab:orbits} pertains to more
distant objects at higher Galactic latitudes, and consequently, greater
distances from the Galactic plane.

The distributions of orbital period and semi-amplitude for our sample 
and the Fekel\,\&\,Hinkle sample are compared in Fig.~\ref{fig:PorbKamp}.  
Our sample peaks at smaller semi-amplitudes and shorter orbital periods. 
Assuming a similar random distribution of orbital inclinations, this
discrepancy could suggest either more massive RGs or lighter WDs in our
sample, or a combination of both. There are, however, important aspects
of sample selection and observational techniques that could artificially 
inflate such a discrepancy.

Our program stars were set from the beginning and the observing campaign
lasted for a maximum of five years for the most observed ones, whereas the
Fekel\,\&\,Hinkle sample was probably drawn from a larger set of objects
observed over 20~yrs.  This naturally favors the predominance of shorter
orbital periods in our sample compared to that of Fekel\,\&\,Hinkle.  It is also
possible that only the systems yielding the cleanest orbits were published
in the Fekel\,\&\,Hinkle sample, which naturally favors larger
semi-amplitudes.

There is an additional possible explanation for the larger
semi-amplitudes of the Fekel\,\&\,Hinkle sample, linked to the adopted
observational technique. Radial velocities for the Fekel\,\&\,Hinkle sample
were derived from molecular CO absorption lines within a narrow interval of
the infrared $H$ band \citep{2000AJ....119.1375F}. The Fekel\,\&\,Hinkle
sample is dominated by burning and HighAcc systems, while ours is entirely
composed of LowAcc objects. The companion in burning and HighAcc SySts is
very hot and luminous, strongly irradiating the facing side of the red
giant. It is plausible that CO is (partially) dissociated by irradiation on
the side of the RG facing the hot WD. This would shift the CO absorption
photocenter toward the shadowed side of the RG (away from the WD), thereby
inflating the orbital semi-amplitude. This irradiation-induced
off-centering might also explain another discrepancy: while all but three of
our 30 orbits are circular, the majority of the 30 Fekel\,\&\,Hinkle orbits
are eccentric.

In conclusion, given these selection and observational biases, the extent to
which the orbits of distant Galah and nearby well-known SySts are
intrinsically different remains unclear.  A better characterization of the
Fekel\,\&\,Hinkle sample's selection function and detailed radiative
modeling of irradiation effects on RVs, combined with the increased Galah
sample size expected in the next paper of this series, will be essential to
resolving this question.

\section{Conclusions}

Following the discovery of SU~Lyn, \citet{2016MNRAS.461L...1M} predicted the
existence of a previously unsuspected and potentially large population of
SySts accreting at much lower levels than those known at
the time.  This series of papers was initiated to test this prediction.  We
began with high-resolution spectra collected for $\sim$0.55 million stars by
the Galah survey up to 2019 \citep{2015MNRAS.449.2604D, 2017MNRAS.464.1259K,
2025PASA...42...51B}, from which we identified $\sim$15k M-type red giants. 
By subtracting median template spectra, we isolated non-pulsating M-type red
giants exhibiting H$\alpha$ emission.  We then subjected a sample of these
candidates to extensive ground-based and space-borne observations to
determine if they hosted accreting degenerate companions.

In Paper~I \citep{2021MNRAS.505.6121M}, we discovered 33 new LowAcc-type
symbiotic stars characterized by accretion luminosities of L$_{\rm acc}$$\leq$10
L$_\odot$ that are similar to SU~Lyn.  In the present work, we shifted our search
toward even lower accretion luminosities ($\leq$3~L$_\odot$) and
successfully discovered, validated, and characterized 44 new symbiotic stars
up to a distance of 5~kpc.  This sample of 77 new SySts significantly
exceeds the total number of previously known, accreting-only Galactic
symbiotic stars and constitutes the bulk of the LowAcc variety discovered to
date.  These objects are of particular interest because it has been argued
that symbiotic binaries spend most of their lifetime in a LowAcc state
\citep{2019arXiv190901389M}.  Properly estimating the total Galactic SySt
population requires accessing the systems currently in these faint phases.

SySts in LowAcc phases are particularly elusive; their UV
excess and X-ray emission typically fall well below the detection thresholds
of current satellites, except for systems very close to the Sun (e.g.,
SU~Lyn at 0.6~kpc).  Their discovery and characterization therefore rely
on optical observations, which must be executed with high precision because
the red giant luminosity outshines the feeble accretion luminosity by a
factor of $\sim$1000.

Between 2021 and 2025, we collected an extensive dataset---spanning over
544 distinct nights---of Echelle spectra using the Asiago 1.82m and Varese
0.84m telescopes.  Combined with the 47 spectra recorded by Galah from
2014--2017, this provides an average of 13.5 visits per each new SySt over a
ten-year baseline.  These high-resolution spectra allowed us to investigate
the time variability of the H$\alpha$ flux and line profiles in detail and
yielded reliable spectroscopic orbits for 22 of the 44 objects, with
preliminary orbits for an additional eight.  An analysis of ASAS-SN photometric
data (2013--2025) revealed light curves modulated by binary effects, such as
irradiation or Roche-lobe filling.  The Gaia XP spectra allowed us to derive
accurate $E_{B-V}$ reddening values, check for chemical peculiarities, and
detect the presence of near-UV excess emission associated with accretion
processes.  Furthermore, high-precision $B$ and $V$ photometry from the
Asiago 67/92cm telescope, with external errors controlled at the 1~mmag
level, allowed us to detect accretion-induced flickering in the majority of
targets (with amplitudes up to 5.7~mmag) and track its stochastic behavior. 
UV and X-ray observations were obtained with the \textit{Swift} satellite
for all 44 new SySts.  The X-ray data excluded the presence of a neutron
star in all targets.  In the UV, three individual stars were detected in the
M2 filter, while the ensemble exhibited a 2.3$\sigma$ excess in
the W2 filter.

The detection thresholds for H$\alpha$ emission and flickering achieved in
this study are unprecedented in the literature.  These capabilities were
instrumental in characterizing these 44 new SySts, which represent the
largest number published in a single paper and more than 10\% of the total
known SySts in the Galaxy and Local Group \citep[see][]{2000A&AS..146..407B,
2019RNAAS...3...28M, 2019ApJS..240...21A}.

Given our precise knowledge of the target-selection function and the Galah
survey footprint, we are confident that upon completion of this project we
will be able to provide an accurate estimate of the total number of SySts in
the Galaxy.  This remains a key ingredient for evaluating their role in
various astrophysical contexts, such as identifying the progenitors of
Type~Ia supernovae or the sources of UV excess in elliptical galaxies.

\section{Data availability}

The full versions of Tables~\ref{tab:logEchelle}, \ref{tab:Ha_evolution},
\ref{tab:flickering}, \ref{tab:Swift}, and \ref{tab:XPf-bins} are only
available in electronic form at the CDS via anonymous ftp to
cdsarc.u-strasbg.fr (130.79.128.5) or via
http://cdsweb.u-strasbg.fr/cgi-bin/qcat?J/A+A/.

\begin{acknowledgements} 
We would like to express our gratitude to the Referee, Dr.  Jennifer
Sokoloski, for her thorough, constructive, and highly insightful report that
helped us clarify several points that were not fully articulated in the
original submission.  This work has been in part supported by INAF 2023
MiniGrant Program (contract C93C23008470001 to UM).  NM acknowledges
financial support through ASI-INAF and ’Mainstream’ agreement 2017-14-H.0
(PI: T.  Belloni).  Part of this work is based on archival data, software or
online services provided by the Space Science Data Center - ASI.  GT
acknowledges financial support of the Slovenian Research Agency (research
core funding No.  P1-0188) and the European Space Agency (Prodex Experiment
Arrangement No.  4000143450).  MP and BJ acknowledge the support of the
Slovenian Research Agency (research core funding No.  P1-0188).  Support by
S.  Dallaporta and F.  Tabacco (ANS Collaboration) is acknowledged. 
\end{acknowledgements}

\bibliographystyle{aa} 
\bibliography{aa60724-26corr} 

\begin{appendix}
\nolinenumbers
\onecolumn

\section{The 44 newly discovered LowAcc symbiotic stars}\label{mainprop}

In the tables of this appendix, we summarize the main properties of the 44
newly discovered LowAcc symbiotic stars, all characterized by accretion at
the lowest detectable levels (L$_{\rm acc}$$\lesssim$3~L$_\odot$), as
thoroughly investigated in this paper.

\begin{table*}[!h]
\caption{Position, brightness and confirmatory indicators: near-UV excess, flickering, orbital motion, and
H$\alpha$ variability.}
\begin{center}
\renewcommand{\arraystretch}{0.84}
\begin{tabular}{cc@{~~}cl@{~~}lc@{~~}c@{~~}cccccccc}
\hline\hline
\multicolumn{1}{c}{(1)}&\multicolumn{1}{c}{(2)}&\multicolumn{1}{c}{(3)}&\multicolumn{1}{c}{(4)}&\multicolumn{1}{c}{(5)}&\multicolumn{1}{c}{(6)}&\multicolumn{1}{c}{(7)}&\multicolumn{1}{c}{(8)}&\multicolumn{1}{c}{(9)}&\multicolumn{1}{c}{(10)}&\multicolumn{1}{c}{(11)}&\multicolumn{1}{c}{(12)}&\multicolumn{1}{c}{(13)}\\
&&&\\ 
name&\multicolumn{2}{c}{RA ~~(J2000)~~DEC}&\multicolumn{1}{c}{$V$}&\multicolumn{1}{c}{$B$$-$$V$}&\multicolumn{3}{c}{Gaia
XP}&Flick.&\multicolumn{1}{c}{RV}&H$\alpha$&\multicolumn{1}{c}{ECH}&A2\\ \cline{6-8}
&&&(mag)&(mag)&$f$-bin&E(B-V)&$\Delta$UV&&orbit&&&int\\
&&&\\       
\hline
&&&\\
GaSS 2$-$34 & 00:45:55.42 &  +01:30:43.9  &  11.092 & 1.673  &   0   & 0.100 & y   & Y!              & Y!    & Y! & 23                 & 001  \\ 
GaSS 2$-$35 & 03:58:21.17 &  +22:48:46.1  &  12.593 & 1.975  &   3   & 0.450 & Y!  & y               & Y!    & Y! & 52                 & 006  \\ 
GaSS 2$-$36 & 04:28:37.67 &  +01:54:15.0  &  12.849 & 1.809  &   6.5 & 0.142 & y:  & Y!              &       & Y! & 19                 & 010  \\ 
GaSS 2$-$37 & 04:34:23.46 &$-$07:32:36.0  &  13.348 & 1.631  &   0   & 0.075 & n   & Y!              &(P!)   & Y! & 12                 & 011  \\ 
GaSS 2$-$38 & 05:00:51.43 &  +16:07:26.6  &  11.857 & 1.911  &   5   & 0.328 & Y!  &                 & Y!    & Y! & 39                 & 023  \\ 
GaSS 2$-$39 & 05:08:03.56 &  +16:37:46.0  &  12.644 & 1.829  &   7.5 & 0.230 & Y   &                 & Y     & Y! & 16                 & 025  \\ 
GaSS 2$-$40 & 05:16:22.01 &  +19:37:00.3  &  11.828 & 2.091  &   6.5 & 0.431 & Y!  & Y!              & Y!    & y  & 28                 & 028  \\ 
GaSS 2$-$41 & 05:26:25.63 &  +05:18:47.1  &  12.582 & 1.840  &   4.5 & 0.130 & n   & y               & y     & Y! & 15                 & 030  \\ 
GaSS 2$-$42 & 08:13:26.66 &$-$04:34:22.3  &  11.125 & 1.700  &   8   & 0.034 & y   & y               &(p)    & Y! & 35                 & 045  \\ 
GaSS 2$-$43 & 08:27:02.89 &  +10:36:03.3  &  12.669 & 1.777  &   2.5 & 0.100 & n   & Y!              & y     & Y! & 16                 & 052  \\ 
GaSS 2$-$44 & 08:27:30.15 &  +13:26:27.1  &  12.067 & 1.606  &   7.5 & 0.010 & ls  & Y               & Y!    & Y! & 35                 & 053  \\ 
GaSS 2$-$45 & 10:32:55.51 &  +12:51:30.4  &  13.557 & 1.494  &   nd  & nd    & nd  & Y!              & Y     & Y! & 7                  & 075  \\ 
GaSS 2$-$46 & 11:11:51.29 &$-$00:20:59.5  &  14.037 & 1.851  &   nd  & nd    & nd  & Y!              & y     & Y! & 7                  & 077  \\ 
GaSS 2$-$47 & 12:08:10.14 &$-$09:07:52.8  &  11.066 & 1.727  &   5.5 &  0.0  & ls  & Y!              &       & Y! & 20                 & 081  \\ 
GaSS 2$-$48 & 13:19:53.78 &$-$07:56:15.6  &  10.392 & 1.673  &   6.5 &  0.0  & n   & Y!              & y     & Y  & 16                 & 085  \\ 
GaSS 2$-$49 & 13:55:15.97 &$-$05:33:49.6  &  11.801 & 1.703  &   5   & 0.020 & ls  & Y!              &       & Y  & 15                 & 091  \\ 
GaSS 2$-$50 & 15:27:23.34 &$-$06:18:39.2  &  12.574 & 1.658  &   7.5 & 0.050 & n   & Y!              &       & Y! & 8                  & 103  \\ 
GaSS 2$-$51 & 15:33:14.40 &$-$16:48:53.2  &  10.760 & 1.627  &   8.5 & 0.130 & Y   & nd              & Y     & y  & 12                 & 106  \\ 
GaSS 2$-$52 & 15:46:20.39 &$-$20:07:30.9  &  12.283 & 1.662  &   8.5 & 0.043 & n   & y               & Y     & nd & 10                 & 122  \\ 
GaSS 2$-$53 & 16:12:59.53 &$-$19:17:23.6  &  13.963 & 1.859  &   3.5 & 0.390 & Y   & nd              &       & nd & 2                  & 160  \\ 
GaSS 2$-$54 & 16:14:15.84 &$-$27:45:19.8 &10.666$^b$& 2.075$^b$& 4.5 & 0.365 & Y!  & nd              & Y!    & nd & 10                 & 164  \\ 
GaSS 2$-$55 & 16:21:39.96 &$-$16:35:53.7  &  12.674 & 1.949  &   0   & 0.430 & Y   & nd              & Y     &    & 5                  & 171  \\ 
GaSS 2$-$56 & 16:26:12.65 &  +06:23:20.3  &  12.801 & 1.727  &   0   & 0.270 & Y!  & Y               & Y!    & Y! & 8                  & 178  \\ 
GaSS 2$-$57 & 16:26:16.23 &$-$27:35:34.3 &12.881$^b$& 2.071$^b$& 4.5 & 0.470 & Y!  & nd              &       & nd & 3                  & 179  \\ 
GaSS 2$-$58 & 16:27:54.21 &  +06:14:41.3  &  12.601 & 1.642  &   8   & 0.050 & n   & Y!              & Y     & Y! & 10                 & 184  \\ 
GaSS 2$-$59 & 16:38:12.61 &$-$28:08:27.0 &13.042$^b$& 1.925$^b$& 4   & 0.408 & Y!  & nd              &(P!)   & nd & 2                  & 206  \\ 
GaSS 2$-$60 & 16:39:03.40 &$-$19:45:13.3  &  11.628 & 1.976  &   8.5 & 0.382 & Y!  & y               & Y     & nd & 5                  & 207  \\ 
GaSS 2$-$61 & 16:56:30.77 &$-$02:40:07.5  &  13.433 & 1.982  &   2.5 & 0.300 & Y   & Y!              &(p)    & y  & 4                  & 216  \\ 
GaSS 2$-$62 & 17:05:03.81 &$-$20:39:54.9  &  13.305 & 1.850  &   1   & 0.345 & Y!  &                 &(P!)   & nd & 4                  & 222  \\ 
GaSS 2$-$63 & 17:07:29.22 &$-$17:52:24.4  &  13.493 & 1.937  &   2   & 0.360 & y   & nd              & y     & nd & 5                  & 224  \\ 
GaSS 2$-$64 & 17:17:26.10 &$-$17:24:51.1  &  11.803 & 1.962  &   3.5 & 0.295 & Y!  &                 & y     & nd & 9                  & 231  \\ 
GaSS 2$-$65 & 17:36:47.58 &$-$08:17:20.3  &  12.395 & 2.243  &   8   & 0.610 & Y!  &                 & y     & y  & 9                  & 234  \\ 
GaSS 2$-$66 & 17:42:16.07 &$-$16:01:50.6  &  12.993 & 1.849  &   7   & 0.295 & Y   &                 &(P!)   & nd & 5                  & 235  \\ 
GaSS 2$-$67 & 17:44:09.21 &  +05:44:58.3  &  12.938 & 1.754  &   7   & 0.290 & Y   & Y!              & Y     & Y! & 10                 & 236  \\ 
GaSS 2$-$68 & 18:11:21.40 &  +06:27:15.8  &  12.989 & 1.798  &   4.5 & 0.180 & y   & $^a$            & Y     & y  & 11                 & 249  \\ 
GaSS 2$-$69 & 18:12:14.13 &  +05:34:47.5  &  12.183 & 1.756  &   5   & 0.170 & Y   &                 & Y     & Y! & 11                 & 250  \\ 
GaSS 2$-$70 & 18:12:45.59 &  +06:00:58.8  &  13.006 & 1.817  &   6.5 & 0.190 & y   &                 & y     & y  & 4                  & 251  \\ 
GaSS 2$-$71 & 18:44:50.54 &$-$22:45:13.0  &  10.427 & 1.892  &   nd  &  nd   & nd  & Y!              & Y     & nd & 7                  & 256  \\ 
GaSS 2$-$72 & 19:01:28.50 &$-$26:48:21.4 &12.523$^b$& 1.913$^b$& 0   & 0.360 & Y!  & nd              &       & nd & 4                  & 292  \\ 
GaSS 2$-$73 & 19:19:53.89 &$-$15:45:22.8  &  10.341 & 1.689  &   5   & 0.055 & Y   & Y               & y     & nd & 9                  & 323  \\ 
GaSS 2$-$74 & 19:43:27.18 &  +02:10:01.3  &  11.443 & 1.915  &  11   & 0.280 & Y!  &                 & Y     & Y  & 15                 & 342  \\ 
GaSS 2$-$75 & 20:09:45.80 &  +00:51:52.7  &  12.331 & 1.689  &   nd  &  nd   & nd  & Y!              & Y     & Y! & 17                 & 354  \\ 
GaSS 2$-$76 & 21:12:29.70 &  +05:39:42.7  &  12.705 & 1.749  &   7   & 0.069 & y:  & Y!              & Y!    & Y! & 15                 & 356  \\ 
GaSS 2$-$77 & 22:13:13.16 &  +07:32:39.8  &  11.641 & 1.687  &   5   & 0.052 & y:  &                 &(P)    & Y! & 21                 & 362  \\ 
&&&\\
\hline 
\end{tabular}   
\end{center}
\tablefoot{The presence of confirmatory indicators -- near-UV excess,
flickering, orbital motion, and H$\alpha$ variability -- is summarized via
qualitative flags.  These indicate presence and increasing prominence or
confidence of the feature, following the sequence 'y', 'Y', and 'Y!'. 
Detailed numerical values and analysis are provided in the relevant tables,
sections, and appendices of this paper.  "nd" indicates no data available. 
Columns: (1) Object name (following Paper~I); (2, 3) RA and Dec from the
Galah~DR3 catalog; (4, 5) median $V$-band magnitude and $B-V$ color from our
2021--2025 photometric monitoring (Asiago Schmidt 67/92cm, except $b$
indicating ANS Collaboration ID 0310 telescope); (6--8) results from fitting
Gaia~XP spectra to $f$-bin templates: best-matching $f$-bin, inferred
$E(B-V)$, and detection of near-UV excess at $\lambda \leq 4000$~\AA\ ("ls"
stands for low signal); (9) $B$-band flickering detection (Asiago Schmidt
67/92cm; $^a$ = a field star 3.6 arcsec away and 1.79~mag fainter prevents
flickering measurements); (10) RV orbit determination (Asiago 1.82m and
Varese 0.84m Echelle spectra); for those lacking one, 'p', 'P', and 'P!'
indicate increasing prominence of an orbital photometric period in ASAS-SN
data; (11) H$\alpha$ flux and profile variability on Asiago 1.82m Echelle
spectra; (12) number of separate nights during which high-resolution Echelle
spectra were collected (Asiago 1.82m and Varese 0.84m); (13) internal
identification number.}
\label{tab:newsymb} 
\end{table*}

\begin{table*}
\caption{Interstellar reddening, distance and accretion luminosity.}
\label{tab:summary} 
\centering
\begin{tabular}{@{}c@{~}c@{~~}cc@{~~}rc@{~~}c@{~~}cccc@{~~}ccl@{}}
\hline\hline
\multicolumn{1}{c}{(1)}&\multicolumn{1}{c}{(2)}&\multicolumn{1}{c}{(3)}&\multicolumn{1}{c}{(4)}&\multicolumn{1}{c}{(5)}&\multicolumn{1}{c}{(6)}&\multicolumn{1}{c}{(7)}&\multicolumn{1}{c}{(8)}&\multicolumn{1}{c}{(9)}&\multicolumn{1}{c}{(10)}&&\multicolumn{1}{c}{(11)}&\multicolumn{1}{c}{(12)}&\multicolumn{1}{c}{(13)}\\
&&&&&&&\\
   name     & $f$-bin &$E_{B-V}$& $B_P$$-$$R_P$ &\multicolumn{1}{c}{$R_P$} & $K_s$ & M2 & dist 
&\multicolumn{2}{c}{H$\alpha$ flux} &&\multicolumn{2}{c}{L$_{\rm acc}$ (L$_\odot$)} 
&\multicolumn{1}{c}{L$_{\rm X}$ (L$_\odot$)} \\ \cline{9-10} \cline{12-13} 
            &      & (mag)  & (mag)  & (mag)   & (mag)   & (mag) &   (kpc) &  mean     & max       && mean &max   & (0.3-10 keV)\\
&&&&&&&\\
\hline
&&&&&&&\\
 GaSS 2-34  &   0  &  0.021 &  1.847 &   9.492 &   7.171 &    19.54 &    1.84 &  2.76e-14 & 6.3e-14  && 0.3  & 0.8  &  $<$0.014  \\  
 GaSS 2-35  &   2  &  0.271 &  2.338 &  10.411 &   7.567 &          &    2.49 &  4.10e-14 & 5.4e-14  && 1.8  & 2.5  &  $<$0.10   \\  
 GaSS 2-36  &   6  &  0.179 &  2.441 &  10.722 &   7.827 &          &    4.07 &  1.24e-14 & 2.1e-14  && 1.2  & 2.1  &  $<$0.052  \\  
 GaSS 2-37  &   0  &  0.063 &  1.811 &  11.777 &   9.494 &          &    5.17 &  5.64e-15 & 1.0e-14  && 0.6  & 1.2  &  $<$0.25   \\  
 GaSS 2-38  &   5  &  0.341 &  2.459 &   9.607 &   6.558 &($>$19.1) &    2.02 &  1.50e-14 & 3.4e-14  && 0.5  & 1.1  &  $<$0.051  \\  
 GaSS 2-39  &   7  &  0.270 &  2.663 &  10.243 &   7.152 &          &    3.20 &  5.17e-15 & 2.3e-14  && 0.3  & 1.7  &  $<$0.093  \\  
 GaSS 2-40  &   7  &  0.492 &  2.742 &   9.263 &   6.035 & $>$21.09 &    1.73 &  5.75e-15 & 2.2e-14  && 0.2  & 0.7  &  $<$0.0070 \\  
 GaSS 2-41  &   5  &  0.190 &  2.206 &  10.660 &   7.921 &          &    3.86 &  6.46e-15 & 1.9e-14  && 0.5  & 1.7  &  $<$0.12   \\  
 GaSS 2-42  &   8  &  0.049 &  2.490 &   8.892 &   6.059 &(20.1)    &    2.21 &  4.85e-14 & 7.6e-14  && 1.0  & 1.7  &  $<$0.035  \\  
 GaSS 2-43  &   3  &  0.037 &  2.012 &  10.863 &   8.403 &($>$19.1) &    4.21 &  2.83e-14 & 4.8e-14  && 2.2  & 3.9  &  $<$0.22   \\  
 GaSS 2-44  &   7  &  0.044 &  2.388 &   9.993 &   7.243 & $>$19.97 &    3.53 &  4.28e-14 & 1.3e-13  && 2.4  & 8.1  &  $<$0.13   \\  
 GaSS 2-45  &   8  &  0.025 &  2.430 &  11.134 &   8.412 & $>$19.67 &    6.44 &  1.01e-14 & 1.7e-14  && 1.8  & 3.1  &  $<$0.74   \\  
 GaSS 2-46  &   3  &  0.032 &  2.093 &  12.198 &   9.760 &          &    7.86 &  5.81e-15 & 8.4e-15  && 1.5  & 2.3  &  $<$0.68   \\  
 GaSS 2-47  &   6  &  0.080 &  2.226 &   9.236 &   6.621 &(20.4)    &    2.32 &  6.30e-14 & 9.7e-14  && 1.6  & 2.6  &  $<$0.30   \\  
 GaSS 2-48  &   7  &  0.035 &  2.261 &   8.393 &   5.736 & $>$19.23 &    1.74 &  3.00e-14 & 9.9e-14  && 0.3  & 1.3  &  $<$0.030  \\  
 GaSS 2-49  &   5  &  0.029 &  2.154 &   9.917 &   7.361 & $>$19.43 &    3.14 &  2.90e-14 & 4.3e-14  && 1.2  & 1.8  &  $<$0.21   \\  
 GaSS 2-50  &   8  &  0.089 &  2.438 &  10.178 &   7.363 &          &    3.91 &  2.28e-14 & 5.1e-14  && 1.7  & 4.1  &  $<$0.23   \\  
 GaSS 2-51  &   9  &  0.098 &  2.776 &   8.199 &   5.219 &          &    1.60 &  6.35e-14 & 1.1e-13  && 0.8  & 1.4  &  $<$0.047  \\  
 GaSS 2-52  &   9  &  0.129 &  2.579 &   9.908 &   6.933 &    19.6  &    3.47 &  3.20e-14 &          && 2.1  &      &  $<$0.26   \\  
 GaSS 2-53  &   2  &  0.409 &  2.292 &  11.959 &   9.020 & $>$19.43 &    4.57 &  7.10e-15 &          && 1.4  &      &  $<$0.24   \\  
 GaSS 2-54  &   4  &  0.436 &  2.469 &   8.484 &   5.401 &    17.76 &    1.01 &  2.00e-13 &          && 2.1  &      &  $<$0.016  \\  
 GaSS 2-55  &   0  &  0.289 &  2.190 &  10.650 &   7.991 & $>$20.36 &    2.47 &  2.53e-14 & 3.1e-14  && 1.1  & 1.4  &  $<$0.073  \\  
 GaSS 2-56  &   0  &  0.062 &  2.021 &  11.034 &   8.615 &          &    3.56 &  2.51e-14 & 3.4e-14  && 1.4  & 2.0  &  $<$0.26   \\  
 GaSS 2-57  &   4  &  0.659 &  2.604 &  10.504 &   7.175 &          &    2.13 &  3.80e-14 &          && 3.0  &      &  $<$0.046  \\  
 GaSS 2-58  &   9  &  0.051 &  2.511 &  10.135 &   7.305 & $>$19.94 &    4.17 &  1.24e-14 & 3.4e-14  && 0.9  & 2.8  &  $<$0.35   \\  
 GaSS 2-59  &   4  &  0.537 &  2.473 &  10.867 &   7.739 & $>$19.13 &    2.84 &  1.30e-14 &          && 1.3  &      &  $<$0.22   \\  
 GaSS 2-60  &   8  &  0.517 &  3.014 &   8.976 &   5.521 & $>$19.07 &    1.50 &  1.52e-14 & 3.0e-14  && 0.4  & 0.8  &  $<$0.078  \\  
 GaSS 2-61  &   3  &  0.249 &  2.295 &  11.512 &   8.684 & $>$20.02 &    4.60 &  2.02e-15 & 6.6e-15  && 0.2  & 0.9  &  $<$0.32   \\  
 GaSS 2-62  &   2  &  0.266 &  2.213 &  11.180 &   8.415 &          &    3.62 &  1.25e-14 & 1.4e-14  && 1.1  & 1.3  &  $<$1.0    \\  
 GaSS 2-63  &   2  &  0.324 &  2.281 &  11.464 &   8.588 & $>$19.39 &    3.89 &  5.33e-15 & 7.9e-15  && 0.6  & 0.9  &  $<$0.62   \\  
 GaSS 2-64  &   4  &  0.217 &  2.283 &   9.734 &   6.952 & $>$20.64 &    2.20 &  5.50e-14 & 6.2e-14  && 1.7  & 1.9  &  $<$0.050  \\  
 GaSS 2-65  &   8  &  0.651 &  3.155 &   9.439 &   5.788 & $>$20.19 &    1.64 &  8.57e-15 & 2.1e-14  && 0.3  & 0.9  &  $<$0.033  \\  
 GaSS 2-66  &   7  &  0.381 &  2.661 &  10.554 &   7.408 &          &    3.42 &  1.72e-14 & 3.1e-14  && 1.9  & 3.5  &  $<$0.92   \\  
 GaSS 2-67  &   9  &  0.184 &  2.684 &  10.478 &   7.448 & $>$20.09 &    4.31 &  3.46e-15 & 1.5e-14  && 0.3  & 1.7  &  $<$0.26   \\  
 GaSS 2-68  &   5  &  0.207 &  2.244 &  10.922 &   8.123 & $>$19.96 &    4.25 &  4.01e-15 & 1.0e-14  && 0.4  & 1.1  &  $<$0.20   \\  
 GaSS 2-69  &   5  &  0.176 &  2.282 &  10.135 &   7.336 &($>$19.1) &    3.01 &  1.69e-14 & 2.8e-14  && 0.9  & 1.5  &  $<$0.17   \\  
 GaSS 2-70  &   7  &  0.187 &  2.478 &  10.737 &   7.866 &($>$19.1) &    4.44 &  3.73e-15 & 6.3e-15  && 0.4  & 0.7  &  $<$0.34   \\  
 GaSS 2-71  &   5  &  0.371 &  2.517 &   8.032 &   4.935 & $>$20.66 &    0.95 &  7.70e-14 & 1.1e-13  && 0.6  & 0.8  &  $<$0.010  \\  
 GaSS 2-72  &   0  &  0.231 &  2.110 &  10.759 &   8.178 & $>$19.29 &    2.73 &  2.70e-14 &          && 1.3  &      &  $<$0.26   \\  
 GaSS 2-73  &   5  &  0.136 &  2.148 &   8.443 &   5.836 & $>$20.23 &    1.48 &  6.88e-14 & 1.2e-13  && 0.8  & 1.4  &  $<$0.021  \\  
 GaSS 2-74  &  10  &  0.303 &  3.341 &   8.285 &   4.821 & $>$20.62 &    1.42 &  1.97e-14 & 4.3e-14  && 0.3  & 0.6  &  $<$0.024  \\  
 GaSS 2-75  &   9  &  0.103 &  2.655 &   9.726 &   6.765 &          &    3.25 &  3.33e-14 & 6.9e-14  && 1.8  & 3.9  &  $<$0.16   \\  
 GaSS 2-76  &   6  &  0.094 &  2.340 &  10.605 &   7.843 & $>$19.58 &    4.18 &  9.96e-15 & 1.9e-14  && 0.8  & 1.6  &  $<$0.38   \\  
 GaSS 2-77  &   5  &  0.052 &  2.152 &   9.740 &   7.128 & (20.5)   &    2.82 &  1.21e-14 & 3.6e-14  && 0.4  & 1.3  &  $<$0.073  \\  
&&&&&&\\
\hline
\end{tabular}
\tablefoot{Columns: (2) $f$-bin spectral classification from the depth of
TiO bands as measured on Galah spectra, following the numerical expressions
in Paper~I; (3) interstellar $E_{B-V}$ reddening from the Galactic dust maps
of \citet{1998ApJ...500..525S}; (4) and (5) Gaia DR3 ($B_P$$-$$R_P$) color
index and $R_P$ magnitude; (6) 2MASS $K_s$ magnitude; (7) UVOT M2 magnitude
as measured by {\it Swift}, summarized from epoch observations listed in
Table~\ref{tab:Swift}, with number in parenthesis providing GALEX NUV values
from Table~\ref{tab:GALEX}; (8) distance computed from $E_{B-V}$, $R_P$, and
$K_s$ against the absolute magnitudes listed in Table~\ref{tab:absmag}; (9)
and (10) mean and max value for the integrated flux (in
erg\,cm$^{-2}$\,s$^{-1}$) of H$\alpha$ emission line from epoch values in
Table~\ref{tab:Ha_evolution} (when only one measurement is available, the
'max' column remains empty); (11) and (12) accretion luminosity (in solar
units) inferred from the H$\alpha$ flux following Eq.(\ref{equat}); (13)
absorbed luminosity in the X-rays over 0.3-10 keV interval from the {\it
Swift} XRT epoch observations listed in Table~\ref{tab:Swift} (see
Sect.~\ref{luminosity}).}
\end{table*}
\clearpage
\newpage
%\twocolumn

\section{Lifetime in the accreting-only phase}\label{lifetime}

By using standard relations, we can estimate the order of magnitude for the
time a symbiotic star spends accreting versus burning. We assume the
accreting WD can begin burning material on its surface without a
mass-expelling flash \citep[as in V4368 Sgr or V1016 Cyg, cf.][figure 6.5
there]{2019arXiv190901389M}, and that the amount of mass transfered during
the burning phase is sufficiently small to be ignored.

Indicating with $\Delta M_{\rm crit}$ the amount of matter a WD 
has to accrete to reach ignition conditions for nuclear burning at the
surface, the time it takes to exhaust the burning is $t_{\rm burn} = \Delta
M_{\rm crit} / \dot{M}_{\rm nucl}$, where $\dot{M}_{\rm nucl}$ is the mass
burning rate.  The time required to accrete the shell is $t_{\rm acc} =
\Delta M_{\rm crit} / \dot{M}_{\rm acc}$, where $\dot{M}_{\rm acc}$ is the
mass accretion rate.  Therefore, the ratio of the time spent accreting the
shell to the time spent burning it is:
\begin{equation}
\frac{t_{\rm acc}}{t_{\rm burn}} = \frac{\dot{M}_{\rm nucl}}{\dot{M}_{\rm acc}}
\end{equation} 
This ratio also represents the expected proportion of SySts in
accreting-only conditions relative to burning systems.

For stable burning conditions, the mass burning rate is governed by the core
mass-luminosity relation, which -- following \citet{1970AcA....20...47P} --
can be expressed as :
\begin{equation}
\left(\frac{L_{\rm burn}}{L_\odot}\right) = 59,250 \left(\frac{M_{\rm WD}}{M_\odot}\right) - 30,950 
\end{equation}
Considering that burning the hydrogen content of 1~gr of solar chemical
mixture generates $4.6 \times 10^{18}$~erg, the mass burning rates for WDs
of 0.6, 1.0, and 1.3~$M_\odot$ are $6.1 \times 10^{-8}$, $3.75 \times
10^{-7}$, and $6.1 \times 10^{-7}$~M$_\odot$\,yr$^{-1}$, respectively.

The mass accretion rate can be expressed as:
\begin{equation}
\dot{M}_{\rm acc} = \frac{L_{\rm acc} R_{\rm WD}}{G M_{\rm WD}}
\end{equation}
and the standard mass-radius relation for WDs is provided by Eggleton's
approximation \citep{1988ApJ...332..193V}:
\begin{equation}
\left(\frac{R_{\rm WD}}{R_\odot}\right) \approx 0.0114 \sqrt{ \left(\frac{M_{\rm WD}}{M_{\rm Ch}}\right)^{-2/3} - \left(\frac{M_{\rm WD}}{M_{\rm Ch}}\right)^{2/3}}
\end{equation} 
where $M_{\rm Ch} \approx 1.44~M_\odot$ is the Chandrasekhar limit.  The
radii for WDs of 0.6, 1.0, and 1.3~$M_\odot$ are 0.013, 0.008, and
0.004~$R_\odot$, respectively.

Accretion luminosities $L_{\rm acc} = 1$ and 10~L$_\odot$ may be considered
representative of SU~Lyn and the new symbiotic stars studied in the present
paper and in Paper~I.  From the expressions above, $L_{\rm acc} = 1~L_\odot$
corresponds to mass accretion rates of $6.88 \times 10^{-10}$, $2.54 \times
10^{-10}$, and $9.78 \times 10^{-11}$~$M_\odot$\,yr$^{-1}$ for WDs of 0.6,
1.0, and 1.3~$M_\odot$, respectively; these rates are a factor of 10 higher
for $L_{\rm acc} = 10~L_\odot$.  The corresponding $t_{\rm acc} / t_{\rm
burn}$ ratios are summarized in Table~\ref{tab:accvsburn}: it is evident
that the burning phase is much shorter than the accretion phase for the
LowAcc SySts discovered in this series of papers.

The values in Table~\ref{tab:accvsburn} assume that the accretion phase
progresses at a constant rate for the entire duration of $t_{\rm acc}$.  For
systems with an accretion disk, this implies that the mass transfer rate to
the disk is constant over time and equal to the mass transfer rate from the
disk to the central WD.  This may not be always the case, tough.  In T~CrB
(in which a well structured accretion disk is fed by Roche-lobe overflow),
the accretion phase does not proceed at a constant rate but cycles through
LowAcc and HighAcc states \citep{2018ApJ...864...46L, 2020ApJ...902L..14L,
2025A&A...701A.176M}.  The LowAcc states appear to correspond to the gradual
buildup of mass in the accretion disk, with little material actually
transferred to the central WD.  Only when the disk grows beyond stable
conditions does it flush most of its content onto the WD in a brighter and
shorter-lived HighAcc phase.  Such a dual accretion regime effectively
increases the time-averaged $\langle L_{\rm acc} \rangle$, thereby
shortening $t_{\rm acc}$ and reducing the $t_{\rm acc}/t_{\rm burn}$ values
reported in Table~\ref{tab:accvsburn}.

 The magnitude of this reduction remains unclear, as it depends on the
buffering properties of the accretion disk.  These properties are generally
unknown for the conditions prevailing in SySts, particularly for WDs
accreting via wind-capture rather than Roche-lobe overflow.

\begin{table}
\caption[]{Ratio $t_{\rm acc}$/$t_{\rm burn}$ of the time spent by a WD
accreting the ignition mass and then burn it non-explosively, as function of
WD mass and accretion luminosity.}
\label{tab:accvsburn}
\renewcommand{\arraystretch}{0.90}
\begin{center}
\begin{tabular}{ccc}
\hline\hline
\noalign{\smallskip}
                    & L$_{\rm acc}$ = 1~L$_\odot$ & L$_{\rm acc}$ = 10~L$_\odot$ \\
\noalign{\smallskip}
\hline
\noalign{\smallskip}
M$_{\rm WD}$ = 0.6~M$_\odot$ &  90 &  9 \\
\noalign{\smallskip}
M$_{\rm WD}$ = 1.0~M$_\odot$ &  1475 &  148 \\
\noalign{\smallskip}
M$_{\rm WD}$ = 1.3~M$_\odot$ &  6235 &  624 \\
\noalign{\smallskip}
\hline
\end{tabular} 
\end{center}
\end{table}  

\section{Recent searches for new symbiotic stars}\label{appendix:literature}

In recent years, substantial effort has been invested in discovering new
symbiotic stars, employing a wide range of approaches.

In addition to a number of more-or-less serendipitous discoveries of
individual new SySts \citep[e.g.,][]{2020AJ....160..125L, 2020A&A...644A..49M,
2022MNRAS.510.1404M, 2025A&A...699A.117M, 2021CoSka..51..103M,
2022A&A...661A.124M, 2025A&A...702A.269M, 2022AstL...48...38M,
2025AstBu..80...58M, 2022ApJ...932..132Y, 2023NewA...9801943P,
2025AstBu..80..620T, 2026arXiv260904036B}, the majority of new
identifications in recent years has been provided by targeted efforts
digging or cross-matching existing stellar catalogs, such as:

\begin{itemize}

\item 16 new SySts were identified by \citet{2008A&A...480..409C,
2010A&A...509A..41C, 2011A&A...529A..56C} and \citet{2014A&A...567A..49R} in
the IPHAS H$\alpha$,$r$,$i$ northern Galactic plane survey and confirmed via
single-visit, low-res spectra (hereafter SiVi-LowRes);

\item investigation of sources listed in the SuperCOSMOS H$\alpha$ survey
led \citet{2013MNRAS.432.3186M}, \citet{2014MNRAS.440.1410M}, and
\citet{2026MNRAS.545f2146M} to discover and SiVi-LowRes confirm 
20, 12, and 14 new SySts, respectively;

\item cross-matching of 2MASS, WISE, IPHAS, VPHAS+, and GALEX catalogs led
\citet{2021MNRAS.502.2513A} and \citet{2023MNRAS.519.6044A} to discover 
and SiVi-LowRes confirm additional 7 new SySts;

\item by working on Galah spectra of Galactic red giants, 33 new SySts were
identified and confirmed by follow-up high-resolution, X-ray/UV, and
flickering observations by \citet[][Paper I]{2021MNRAS.505.6121M};

\item by digging the LAMOST spectral database, \citet{2023RAA....23j5012J},
\citet{2025ApJ...995...14Z}, and \citet{2025ApJ...987..147C} identified a
total of 9 new SySts;

\item \citet{2025MNRAS.542.3097B}
discovered and SiVi-LowRes confirmed two new SySts among IPHAS and UVEX
footprint;

\item \citet{2025MNRAS.543.2292L}
investigated outliers in the SkyMapper photometric catalog and their rapid
variability among the three 20-min consecutive exposures collected for the
$u$$v$$g$ bands, and found 12 new SySts confirmed by SiVi-LowRes, X-rays and
flickering observations.
\end{itemize}

In addition to these validated new SySts, most of the above papers also
proposed a number of candidate SySts for later scrutinity, while other
papers limited to compile lists of possible new SySts without subjecting
them to confirmatory follow-up observations, like the 72 objects isolated by
\citet{2019MNRAS.483.5077A} from 2MASS and WISE catalogs, the 25 ones found
by \citet{2025OJAp....8E.122B} in the cross-match of Gaia DR3 products with
X-rays and infrared surveys, the 10 sources found by
\citet{2024ApJ...962..126X} via a joint analysis of GALEX, XMM, and Gaia DR3
catalogs, or the 12 objects proposed by \citet{2026A&A...708A..28C} from
combined analysis of Gaia, 2MASS, WISE, and SkyMapper data.

\section{Absolute magnitudes and intrinsic colors for f-bin spectral
standards}\label{appendix:absmag}

\begin{table*}
\caption{Absolute magnitudes and intrisic colors for the $f$-bin
classification system of M giants introduced in Paper~I.}
\renewcommand{\arraystretch}{0.95}
\begin{tabular}{@{}cccccccccccccc@{}}
\hline\hline
&&&\\
               &&\multicolumn{4}{c}{absolute magnitudes}&&\multicolumn{7}{c}{intrinsic colors}\\ \cline{3-6}\cline{8-14}
$f$-bin & Num. & M($V$) & M($R$$_P$) & M($K$$_s$) & M($W2$) && $B$$-$$V$ & $B$$_P$$-$$R$$_P$ & $J$$-$$K_s$ & $H$$-$$K_s$ & $W1$$-$$W2$ & $W2$$-$$W3$ & $W2$$-$$W4$ \\
&&&\\
\hline
&&&\\
0 & 984  & $-$0.591 & $-$1.947 & $-$4.078 & $-$4.02 && 1.511 & 1.661 & 0.939 & 0.192 & $-$0.14 &   0.15  & 0.27 \\  
1 & 710  & $-$0.672 & $-$2.089 & $-$4.279 & $-$4.22 && 1.536 & 1.716 & 0.963 & 0.204 & $-$0.15 &   0.15  & 0.25 \\
2 & 594  & $-$0.721 & $-$2.215 & $-$4.462 & $-$4.40 && 1.563 & 1.778 & 0.982 & 0.213 & $-$0.16 &   0.15  & 0.25 \\
3 & 478  & $-$0.834 & $-$2.381 & $-$4.695 & $-$4.63 && 1.585 & 1.846 & 1.008 & 0.221 & $-$0.16 &   0.16  & 0.27 \\
4 & 423  & $-$0.873 & $-$2.463 & $-$4.839 & $-$4.78 && 1.609 & 1.919 & 1.028 & 0.230 & $-$0.15 &   0.16  & 0.28 \\
5 & 396  & $-$0.944 & $-$2.656 & $-$5.117 & $-$5.07 && 1.616 & 2.005 & 1.053 & 0.243 & $-$0.14 &   0.16  & 0.28 \\
6 & 341  & $-$0.963 & $-$2.735 & $-$5.272 & $-$5.20 && 1.626 & 2.095 & 1.069 & 0.254 & $-$0.13 &   0.17  & 0.29 \\
7 & 283  & $-$0.958 & $-$2.878 & $-$5.481 & $-$5.44 && 1.628 & 2.215 & 1.094 & 0.262 & $-$0.12 &   0.17  & 0.30 \\
8 & 254  & $-$0.887 & $-$2.957 & $-$5.654 & $-$5.64 && 1.623 & 2.352 & 1.114 & 0.276 & $-$0.10 &   0.17  & 0.31 \\
9 & 238  & $-$0.766 & $-$3.035 & $-$5.859 & $-$5.85 && 1.599 & 2.556 & 1.138 & 0.294 & $-$0.08 &   0.17  & 0.32 \\
10& 190  & $-$0.569 & $-$3.101 & $-$6.106 & $-$6.13 && 1.582 & 2.840 & 1.173 & 0.313 & $-$0.02 &   0.19  & 0.33 \\
11&  96  & $-$0.309 & $-$3.166 & $-$6.376 & $-$6.47 && 1.570 & 3.168 & 1.202 & 0.335 & ~~~0.04 &   0.21  & 0.35 \\
12&  50  & $-$0.005 & $-$3.103 & $-$6.468 & $-$6.61 && 1.561 & 3.348 & 1.227 & 0.352 & ~~~0.06 &   0.27  & 0.38 \\
&&&\\
\hline 
\end{tabular}   
\tablefoot{$B$ and $V$ bands are on the \citet{2009AJ....137.4186L} system, as
realized all-sky by the APASS survey, $B_P$ and $R_P$ are from Gaia
DR3, $J$$H$$K_s$ from 2MASS, and W1--W4 from AllWISE (the latter quoted to
two decimal places due to lower intrinsic photometric accuracy).  The error
on the quoted values is approximately 0.001 mag for APASS, Gaia, and
2MASS data (except for $f$-bin=11, where it is 0.003 mag, rising to 0.005 mag
for $f$-bin=12) and $<$0.01 mag for AllWISE data (0.02 mag for $f$-bin=12).}
\label{tab:absmag}
\end{table*} 

The definition of the 13 $f$-bins introduced in Paper~I -- based on the
depth of TiO band-heads measured from Galah high-resolution spectra -- was
adopted to overcome the inherent limitations of the classical M0--M10\,III
classification system.  While widely utilized, the M0--M10\,III sequence
lacks the sharp, unambiguous definitions required for high-resolution
spectroscopy and suffers from a lack of portability to red wavelengths.

In fact, the original MKK system (comprising the M0--M10\,III sequence) was
established using low-resolution, blue-range (3800--4900~\AA), prismatic
spectra, where the dispersion is strongly dependent on wavelength.  This
system was inherently dependent on the specific dispersion used (quoted as
\AA/mm at H$\gamma$), as it was not based on numerical classification
criteria (such as intensity ratios of selected absorption lines or molecular
bands).  Instead, it required "a predefined set of standard stars to be
reobserved with the given instrumentation and look for the best overall
match with the spectrum of the star under investigation"
\citep{1943assw.book.....M}.  The concept of "best overall match" remained
unchanged in later refinements \citep[e.g.,][]{1976aasc.book.....K} and
during the transition to digital or grating spectra at longer wavelengths
\citep[e.g.,][]{1987clst.book.....J}, with the definition of diagnostic
line or band ratios remaining qualitative rather than quantitative
\citep[e.g.,][]{1977aars.book.....Y}.

To further complicate the application of the M0--M10\,III sequence to Galah
spectra, several conflicting versions of the sequence exist that are
mutually incompatible \citep[e.g.,][]{1987clst.book.....J,
1994A&AS..105..311F}.  For instance, an M2\,III type in the Mt.\ Wilson
system \citep{1953ApJ...117..313J, 1962ApJ...136...21M} corresponds to
M0.6\,III in the Case Western Reserve version \citep{1949ApJ...109..391N,
1964ApJ...139..190N}.  Such discrepancies have significant ramifications for
the application to all-sky spectral surveys and homogeneity across both
hemispheres \citep[e.g.,][]{1964cbs..book.....H, 1975mcts.book.....H},
to spectral atlases \citep[e.g.,][]{1970aops.book.....S, 1977aars.book.....Y},
and the calibration of absolute magnitudes and intrinsic colors
\citep[e.g.,][]{1987ApJ...320..199F, 2007AJ....134.1089S,
2009BaltA..18...19S}.

Calibrating the 13 $f$-bins in terms of absolute magnitudes and
reddening-free intrinsic color indices is therefore of primary interest.  To
this end, we selected M-type giants from the Galah 32k sample (see
Sect.~\ref{XPspectra}) that satisfy three criteria: a Gaia~DR3 parallax
error of less than 10\%, an interstellar reddening upper limit of
$E_{B-V}$$\leq$0.1~mag \citep{1998ApJ...500..525S}, and absence of H$\alpha$
emission even at the weakest detectable level on the template-subtracted
spectra.  This calibration was not feasible in Paper~I because the initial
M-giant sample was smaller (15k versus the current 32k) and, crucially,
Gaia~DR2 parallaxes were significantly less accurate than the current DR3
values for distant objects.

   \begin{figure*}
   \centering
   \includegraphics[angle=270,width=17.8cm]{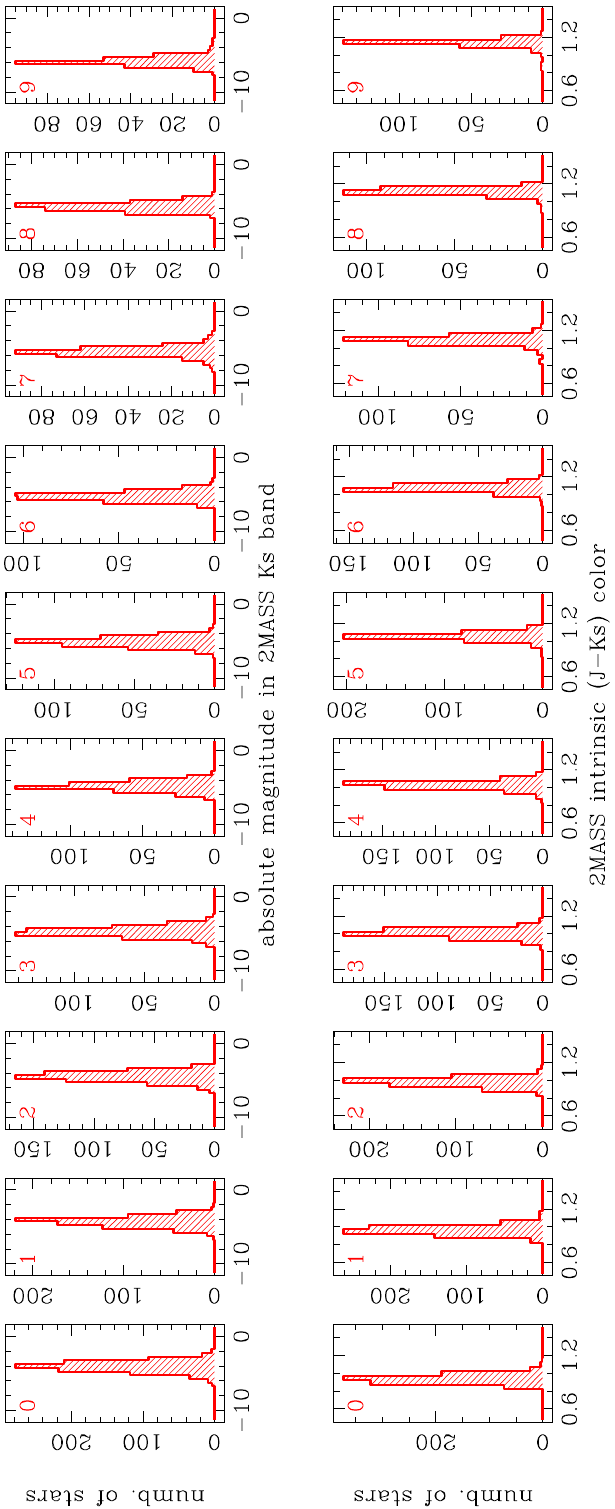}
      \caption{Examples of distribution in absolute magnitude and intrinsic
               colors of the 5,040 M-type red giants selected to calibrate
               the $f$-bin sequence in Table~\ref{tab:absmag}, highlighting
               Gaussian symmetric shapes and lack of outliers.  The $f$-bin
               value is plotted in red.}
         \label{fig:fbinprogression}
   \end{figure*}

   \begin{figure*}[ht!]
   \centering
   \includegraphics[angle=270,width=17.3cm]{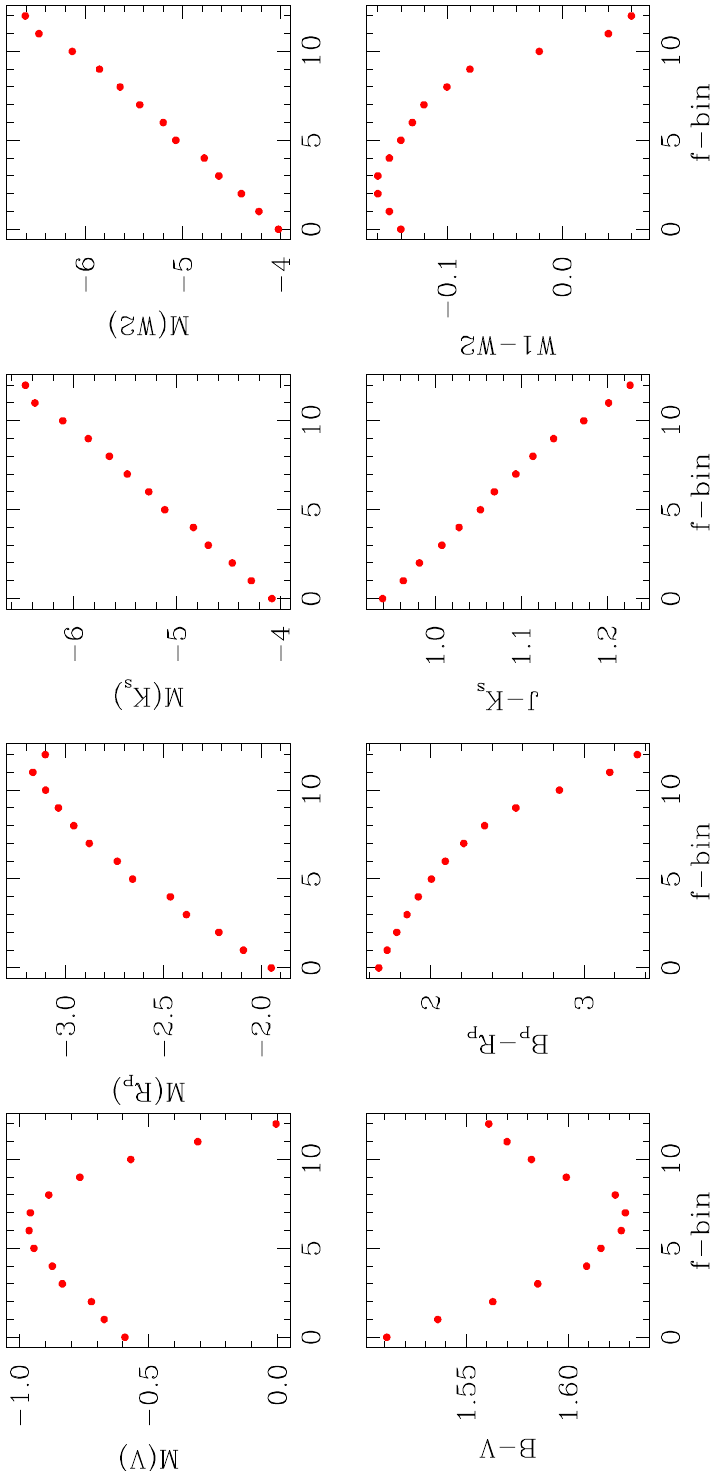}
      \caption{Progression along the $f$-bin sequence of the absolute 
               magnitudes and intrinsic colors listed in Table~\ref{tab:absmag}.}
         \label{fig:fbincalib}
   \end{figure*}

A total of 5,040 M-giants satisfied these parallax and reddening cuts. 
Their distribution across the $f$-bins is reported in
Table~\ref{tab:absmag}, which lists the mean absolute magnitudes and
intrinsic colors computed from the 2MASS \citep{2003yCat.2246....0C}, APASS
DR8 \citep{2014CoSka..43..518H,2015AAS...22533616H}, and AllWISE
\citep{2014yCat.2328....0C} catalogs, in addition to Gaia $B_P$ 
and $R_P$.  Reddening corrections for 2MASS $J$$H$$K_s$ and
APASS DR8 $B$$V$ data were applied using the temperature-dependent relations
from \citet{2003A&A...401..781F}.  Corrections for Gaia $B_P$ and $R_P$
followed the polynomial expressions from Gaia
GSP-Phot \citep{2023A&A...674A..27A}, while for the AllWISE W1, W2, W3, and
W4 bands, we adopted the temperature- and extinction-dependent coefficients
from \citet{2023ApJS..264...14Z}.  The distributions around the mean values
for the 5,040 M-giants are remarkably sharp and Gaussian, showing no
significant outliers, as illustrated by the sample distributions in
Fig.~\ref{fig:fbinprogression}.  The progression of absolute
magnitudes and intrinsic colors along the $f$-bin sequence is smooth and
gradual as shown by Fig.~\ref{fig:fbincalib}.

\section{High-resolution Echelle spectroscopy}\label{appendix:Echelle}

The 1.82m Copernico telescope, located in Asiago (elevation 1370m,
Venetian Prealps), is operated by the National Institute for Astrophysics
(INAF, Italy) through the Astronomical Observatory of Padova.  The
3550--7100 \AA\ interval is covered by the REOSC Echelle spectrograph in 32
orders without inter-order gaps, using an Andor DW436-BV camera (equipped
with an E2V CCD42-40 AIMO CCD, 2048$\times$2048 pixels, 13.5~$\mu$m pixel size). 
The resolving power is $R$$\sim$22,000 for a 1.8-arcsec slit width.  The slit is
typically rotated to the parallactic angle to compensate for atmospheric
chromatic dispersion when observing at low elevations.

Echelle spectra were also collected with the Varese 0.84m telescope,
operated by the ANS Collaboration at the Schiaparelli Observatory in Campo
dei Fiori (elevation 1225m, Varese Prealps).  This facility is equipped with
a Mark III Multi-Mode Spectrograph (MMS) from Astrolight Instr.  The camera
at the Echelle focus of the MMS is an SBIG ST10XME CCD; the recorded
4250--8850 \AA\ range is covered in 27 orders without inter-order gaps.  A
1$\times$1 binning and a 1.8-arcsec slit width provide a resolving power of
$R$$\sim$17,000, which decreases to $\sim$10,000 when using a 2.0-arcsec
slit and 2$\times$2 binning.

Data reduction for both telescopes was performed using IRAF following
standard procedures, including bias, dark, and dome-flat field corrections. 
The slit height of 22 arcsec on both instruments allows for 8 arcsec of
clear sky on either side of the stellar spectrum.  This ensures an accurate
definition and subtraction of the sky background, which is particularly
important during moonlit nights given the faintness of many program stars. 
Wavelength calibration was achieved by obtaining Thorium-Argon lamp
exposures before and after each science frame.  Flux calibration was
performed using spectrophotometric standards observed throughout the night,
followed by heliocentric velocity correction.  Finally, the individual
Echelle orders were merged into a single one-dimensional spectrum, provided
both as a flux-calibrated version (primarily for investigating H$\alpha$
emission) and a continuum-normalized version (adopted for radial velocity
measurements). The zero-point of the fluxed version was checked (and if
necessary corrected) againts the near-simultaneous $g$-band photometry
provided by the ASAS-SN sky patrol \citep{2014ApJ...788...48S,
2023arXiv230403791H}.

\section{Photomeric observations in search of flickering}\label{appendix:flickering}

\subsection{Observing mode and data reduction}

A typical flickering run acquired for this paper lasted 70min, with
observations in the $B$ band interspersed with those in $V$ in a
$V$$B$$B\cdots B$$B$$V$ ---  $V$$B$$B\cdots B$$B$$V$ pattern.  The
maximum number of consecutive $B$-band exposures was 10, provided the time
interval between successive $V$-band exposures did not exceed 10min.  The
exposures in $V$ were required to build the ($B$-$V$) color base to construct
and solve the color equations necessary to transform each individual data
point from the local instantaneous photometric system to the
\citet{2009AJ....137.4186L} standard one.  In solving the color equations,
the $V$ frame closest in time to the given $B$ frame was chosen.  The color
equations are as usual: $B = b + \alpha_b \times (b-v) + \gamma_b$, $ V = v
+ \alpha_v \times (v-i) + \gamma_v$, and $B-V = \beta_{bv} \times (b-v) +
\delta_{bv}$, where $b$,$v$ denote magnitudes in the instantaneous local
system, and $B$,$V$ those in the Landolt standard system.  The color
$B$$-$$V$ is derived directly and not computed as the difference of $B$ and
$V$ values.

Aperture photometry was carried out for all targets, with the aperture
radius set to 1.0$\times$FWHM, and the inner and outer radii of the sky
annulus set at 2.5$\times$ and 3.5$\times$FWHM, respectively.  The target
GaSS 2-68 has an optical companion fainter by $\Delta B$=1.79mag at 3.65
arcsec distance, which spoils the accuracy of the aperture photometry.  We
measured the images of GaSS 2-68 via PSF-fitting, but the limited accuracy
of the method resulted in RMSE=0.011 (cf.  Eq.(\ref{bor})), too large to
allow detecting any flickering with an amplitude similar to that of the
other program stars.

In order to achieve a total error budget as small as 0.002--0.003~mag on the
target and the check stars, the Poissonian contribution cannot exceed
0.001~mag; this requires adjusting the $B$-band exposure time to the target
brightness, ranging from 10sec for the brightest targets to 480sec for the
faintest ones.  It is therefore clear that the presence of flickering is
probed over very different time-scales depending on the $B$-band brightness
of the target.

The flickering light curves secured for the present study are similar
to those presented in Paper 1, with its figure~10 showing illustrative examples,
where the effect of the observational cadence and the interleaving of $B$- and
$V$-band images can be readily appreciated.

\subsection{Comparison sequences}

A total error budget as small as 0.002--0.003~mag for the whole 70min run
requires the definition of a well-chosen and richly populated local $B$,$V$
photometric sequence (LPS), as the all-sky photometric approach is unable to
guarantee such protracted precision.  The LPSs for all program stars were
selected from data release DR8 of the APASS all-sky $B$$V$$g$$r$$i$ survey
\citep[cf.][]{2014CoSka..43..518H}, and composed of at least 50 non-variable
stars widely distributed in color and generally $\sim$1~mag brighter than
the program star.  A preliminary observation of each target checked if any
of the stars selected for the LPS deviated by $\geq$3$\sigma$ from the
linear color relations; such stars were subsequently eliminated from the
LPS.  Once a pruned LPS was established, it remained fixed for all visits to
the given target.

   \begin{figure}
   \centering
   \includegraphics[width=8.8cm]{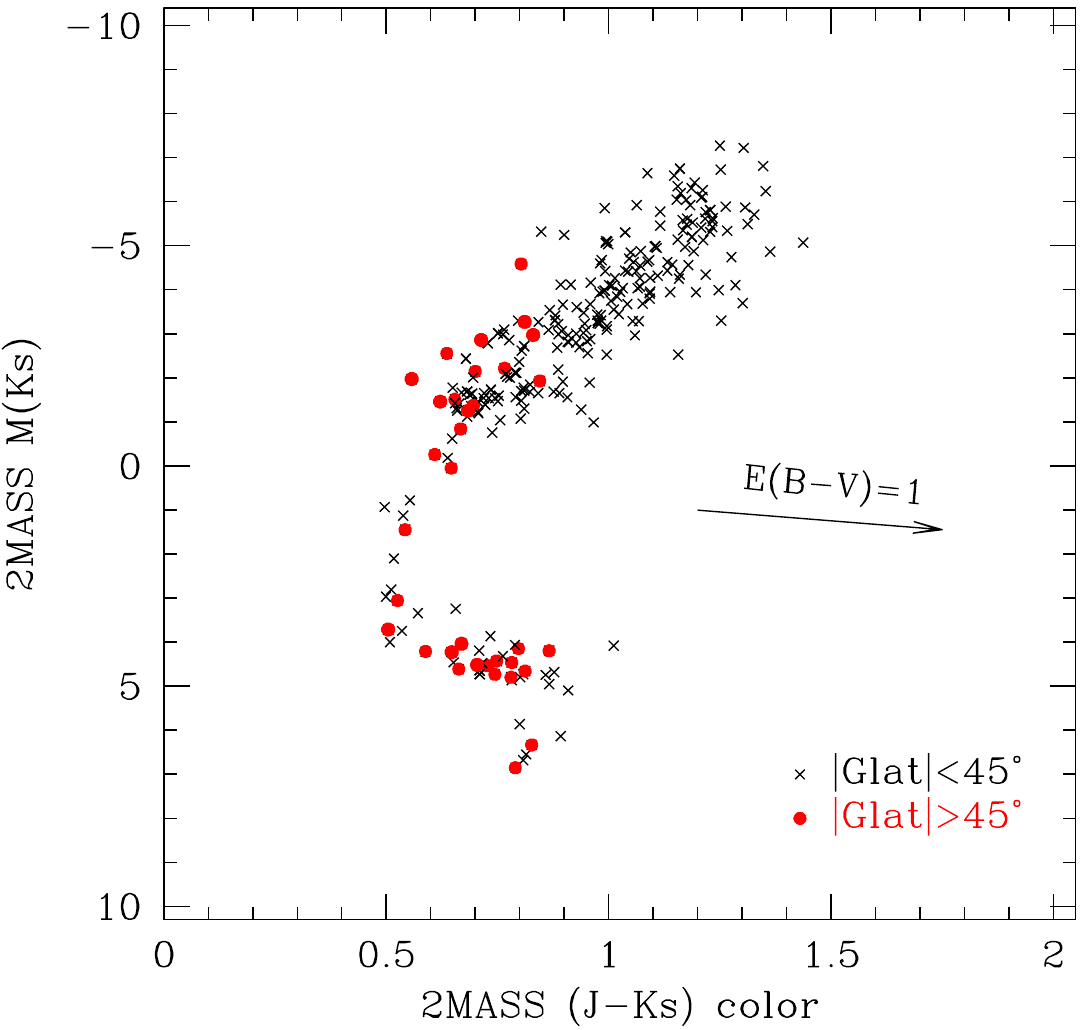}
      \caption{Distribution of the flickering check stars on the
               Hertzsprung-Russell diagram, based on 2MASS $JHK_s$ data and
               Gaia DR3 parallaxes. Different symbols are used to distinguish
               lower from higher Galactic latitude check stars. The reddening
               vector is based on the \citet{1999PASP..111...63F} extinction 
               law for the $R_V = 3.1$ case.}
         \label{fig:checkstars}
   \end{figure}

\subsection{Check stars and the red giants among them}\label{RGflick}

To establish the presence of flickering, the dispersion $\sigma (B)$ of the
measurements during the 70min visit to a target must be compared to the
$\sigma$ of suitable check stars (CKSs), which are not part of the LPS to
avoid circular arguments.  To attain the desired accuracy, the color of the
CKSs has to be close to that of the target (i.e., very red) and they need to
be well distributed in magnitude around that of the target to properly
define the shape and breadth of the Poissonian relation (the gray band in
Fig.~\ref{fig:flickering}).  The choice of the CKSs is made for each target
by selecting those with the most suitable $B$ and $B$$-$$V$ values among
field stars, as measured during the preliminary observation used to prune
the LPS. 

The CKS sample comprises both red dwarfs and red giants, as illustrated
in Fig.~\ref{fig:checkstars}.  Here, all CKSs with an accurate Gaia DR3
parallax (fractional error less than 20\%) are plotted in an infrared
version of the Hertzsprung-Russell diagram based on 2MASS $J$$H$$K_s$ data. 
This choice minimizes the impact of interstellar reddening, which is
inherently low given that the Galah program stars are located away from the
Galactic plane.  As expected from the fixed range in magnitude of Galah
stars (12$\leq$$V$$\leq$14 mag) and the spatial distribution of Galactic
stellar populations \citep[e.g.,][]{2008AJ....136..421Z}, the highest
fraction of red dwarfs among the CKSs is observed at high Galactic
latitudes, whereas red giants dominate at lower Galactic latitudes as
clearly shown in Fig.~\ref{fig:checkstars}.

The fact that the giants within our CKS sample cleanly populate the entirety
of the red giant branch, and yet follow the same Poissonian relation as the
red dwarfs (the gray band in Fig.~\ref{fig:flickering}), demonstrates that
the flickering observed in the program stars is not an intrinsic property of
red giants (either RGB or AGB), but is instead directly associated with
symbiotic activity.

\section{Template Gaia XP spectra mapping the $f$-bin classification
system}\label{appendix:XPspectra}

   \begin{figure*}
   \centering
   \includegraphics[width=0.4633\textwidth]{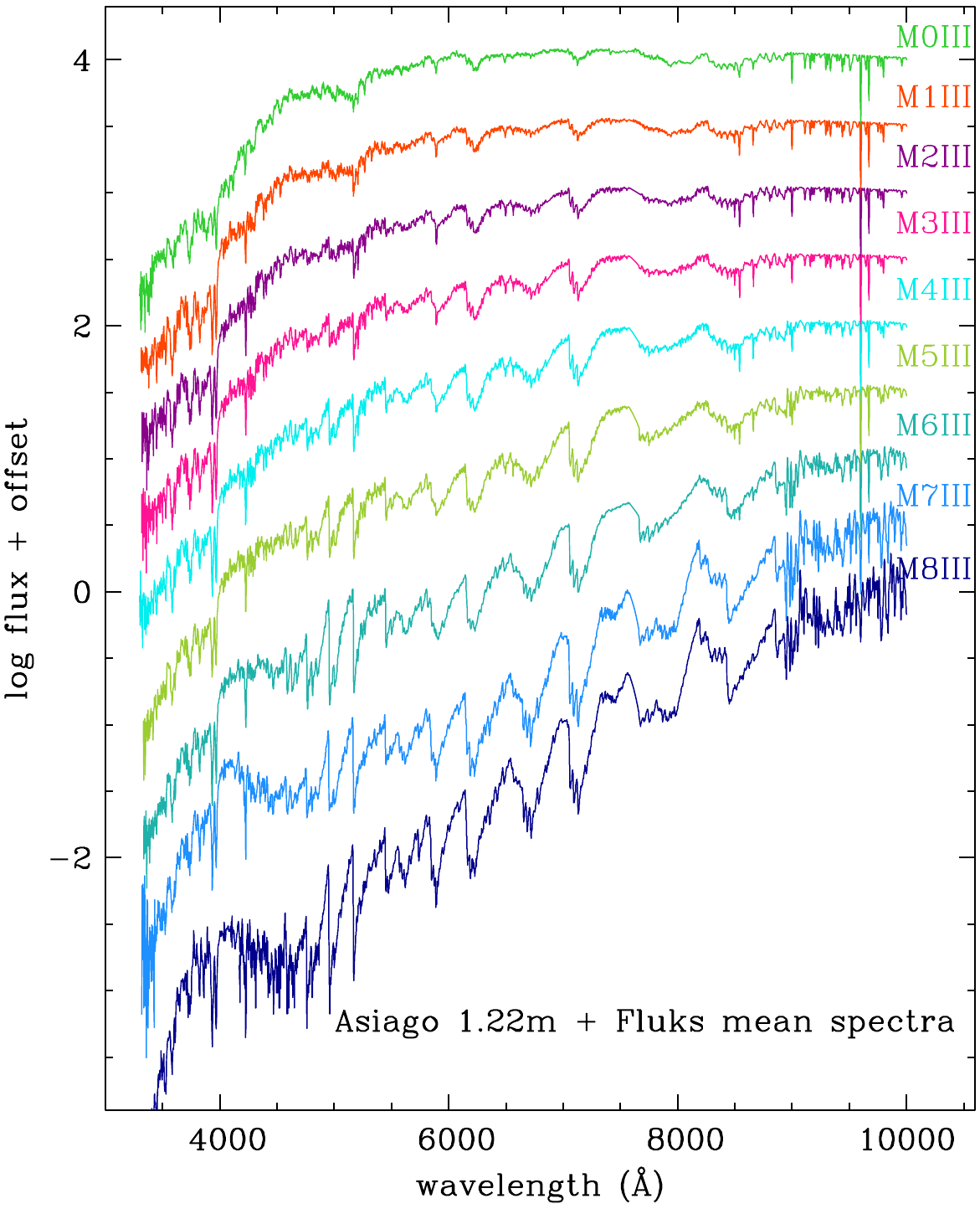}
   \includegraphics[width=0.4167\textwidth]{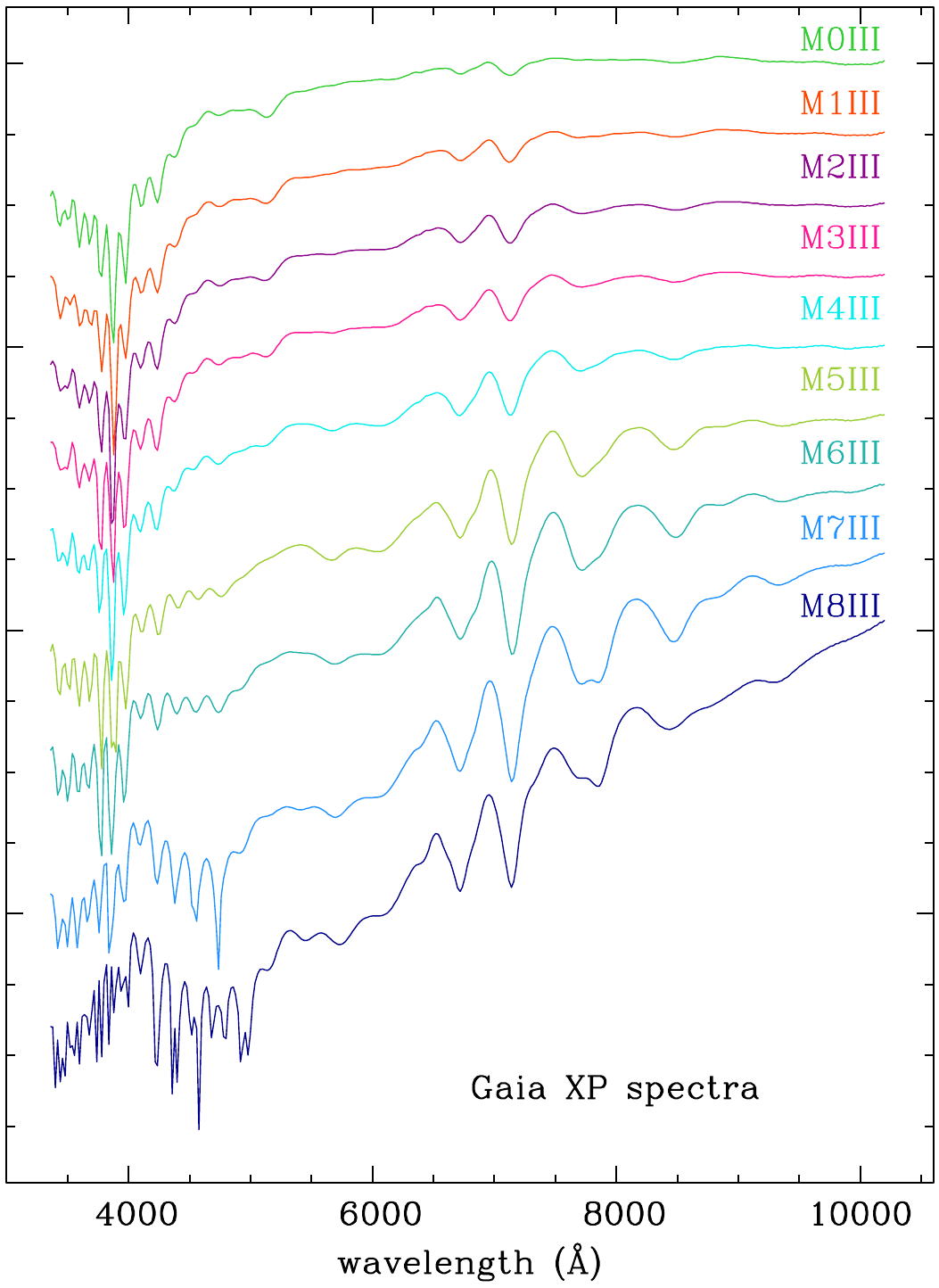}
        \caption{Comparison between classical optical spectra of M giants
        and their Gaia XP counterparts.  The left panel shows mean 3300–8000
        Å spectra obtained with the Asiago 1.22m telescope, extended to
        10,000 Å using data from \citet{1994A&AS..105..311F}.  The right
        panel displays the corresponding Gaia XP mean spectra for the same
        set of stars.}
        \label{fig:XPsymbio}
   \end{figure*}

During the final stages of the Gaia payload design, the originally proposed
14-band photometric system \citep{2006MNRAS.367..290J} was replaced by
low-resolution, slitless prism spectra.  These consist of a blue arm ($B_P$,
330–680 nm) and a red arm ($R_P$, 640–1050 nm), with minimal overlap near
H$\alpha$.  The nominal resolution R ranges from 20 to 100 for $B_P$, and
from 70 to 100 for $R_P$.  Rather than standard flux tables, Gaia spectra
are stored as coefficients of a linear combination of Hermite functions (55
for each arm) to optimize storage and computation.  For Gaia DR3, these
coefficients were converted into sampled XP spectra (flux in W m$^{-2}$
nm$^{-1}$ vs.  wavelength in nm), provided at 10 nm steps from 336 to 1020
nm.  These mean spectra, based on the first 34 months of operations, are
available for $\sim$220 million sources with $G$$\leq$17.65 mag.  Detailed
descriptions and validation are provided in \citet{2021A&A...652A..86C},
\citet{2023A&A...674A...2D}, \citet{2023A&A...674A...3M} and
\citet{2023A&A...674A..32B}.

Two aspects of XP spectra are particularly relevant for SySt studies: ($i$)
the slitless prism dispersion strongly compresses the data toward longer
wavelengths (especially in the $B_P$ arm), making the spectral resolution
highly wavelength-dependent, and ($ii$) the Hermite function representation
-- while efficient for smooth stellar continua -- struggles to reproduce
sharp features such as strong emission lines or "sawtooth" molecular bands
(e.g., the TiO $\alpha$-system bandheads at 4761, 4955, and 5167 \AA). 
Consequently, Gaia XP spectra differ significantly from standard optical
spectra, as shown for the M0–M8III sequence in Fig.~\ref{fig:XPsymbio}. 
Therefore, Gaia XP spectra of SySts must be evaluated against templates
built within the same XP framework.

\begin{figure*}[ht!]
   \centering
   \includegraphics[width=15.5cm]{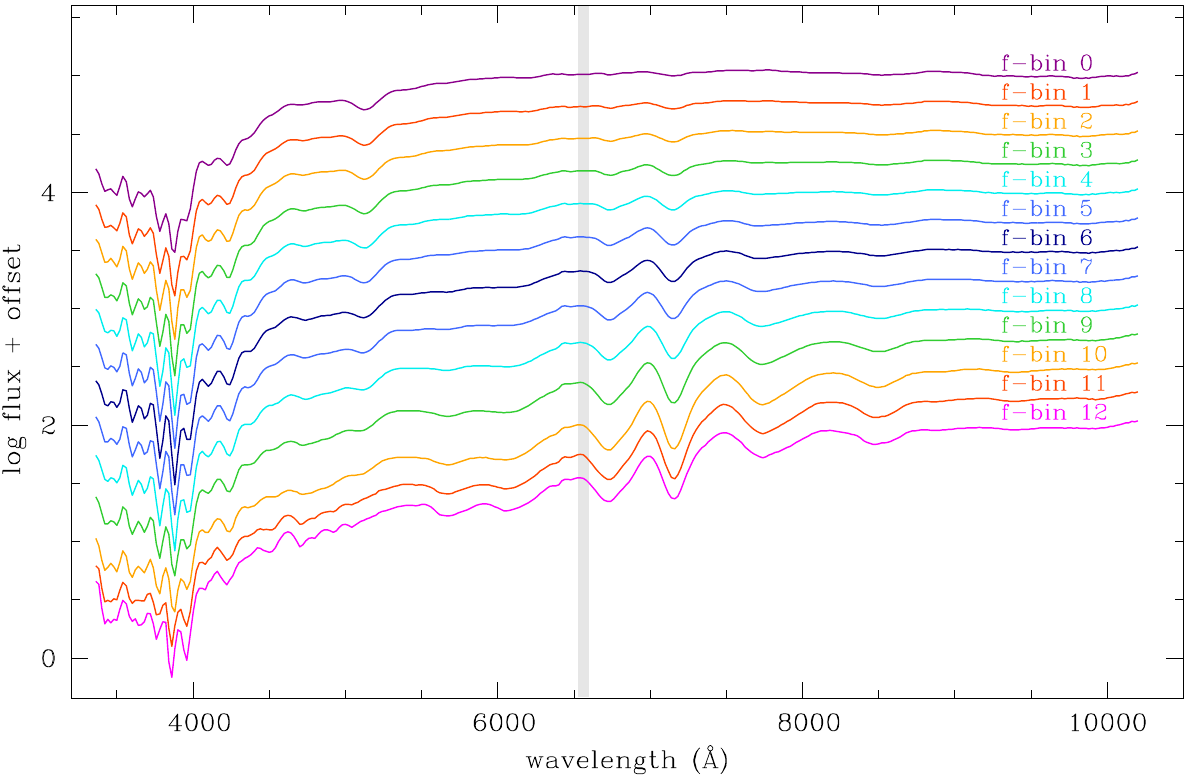}
      \caption{Reddening-free Gaia XP template spectra mapping the $f$-bin 
               classification system. The gray line marks the 
               position of H$\alpha$.}
         \label{fig:XPf-bins}
   \end{figure*}

To construct our templates, we retrieved XP spectra for the 32k sample Galah
M-giants defined in Sect~\ref{XPspectra}.  We retained only sources with low
reddening (E$_{B-V}$$\leq$0.1), high S/N, and no quality flags (about 500
each $f$-bin).  After correcting for interstellar reddening using
\citet{1998ApJ...500..525S} maps and the standard $R_V$=3.1 extiction law by
\citet{1999PASP..111...63F}, we medium combined the spectra into mean
templates for each $f$-bin.  The resulting 13 template spectra are shown in
Fig.~\ref{fig:XPf-bins} and tabulated in Table~\ref{tab:XPf-bins}.  
They exhibit remarkably smooth transitions between $f$-bins, preserving
subtle features at blue wavelengths that might otherwise be mistaken for
noise upon cursory inspection.  The impact of individual anomalies among the
source stars is removed by the median-combination of approximately $\sim$500
individual spectra into each template.  Consequently, any absorption
features typical of S-type giants (whose frequency in the field is $\sim$1\%
of that of normal M-type giants) are automatically filtered out, as is the
presence of emission lines (which may occur in the small fraction of input
giants that are Mira pulsating variables, of the order of a few percent of
the total according to \citet{1996A&ARv...7...97H}) or the signature of
companions (primarely main-sequence stars).

\begin{table*}[ht!]
\centering
\caption{Reddening-free Gaia XP template spectra for the f-bin classification system}
\begin{tabular}{ccccccccc}
\hline\hline
&&&\\
$\lambda$(\AA)             &\multicolumn{8}{c}{log flux (erg cm$^{-2}$ s$^{-1}$ \AA$^{-1}$)}\\ \cline{2-9} 
            & $f$-0       & $f$-1       & $f$-2       & $f$-3        & $f$-4        & $f$-5        & $f$-6       & $f$-7        \\ 
&&&\\
\hline
&&&\\
      10,200 & $-$14.52378 & $-$14.49784 & $-$14.47881 & $-$14.45454  & $-$14.42847  & $-$14.39924  & $-$14.36778 & $-$14.33206 \\ 
      10,180 & $-$14.53819 & $-$14.51207 & $-$14.49217 & $-$14.46784  & $-$14.44060  & $-$14.41209  & $-$14.37972 & $-$14.34315 \\ 
      10,160 & $-$14.53943 & $-$14.51288 & $-$14.49213 & $-$14.46828  & $-$14.44113  & $-$14.41154  & $-$14.37821 & $-$14.34210 \\ 
      10,140 & $-$14.55275 & $-$14.52525 & $-$14.50488 & $-$14.48021  & $-$14.45388  & $-$14.42401  & $-$14.38964 & $-$14.35441 \\ 
      10,120 & $-$14.55215 & $-$14.52438 & $-$14.50430 & $-$14.47896  & $-$14.45343  & $-$14.42336  & $-$14.38902 & $-$14.35524 \\ 
      10,100 & $-$14.55647 & $-$14.52956 & $-$14.50961 & $-$14.48412  & $-$14.45902  & $-$14.42974  & $-$14.39513 & $-$14.36277 \\ 
      10,080 & $-$14.55812 & $-$14.53196 & $-$14.51187 & $-$14.48719  & $-$14.46222  & $-$14.43297  & $-$14.39914 & $-$14.36655 \\ 
      10,060 & $-$14.55376 & $-$14.52878 & $-$14.50821 & $-$14.48373  & $-$14.45821  & $-$14.43052  & $-$14.39743 & $-$14.36534 \\ 
      10,040 & $-$14.55684 & $-$14.53181 & $-$14.51184 & $-$14.48703  & $-$14.46226  & $-$14.43534  & $-$14.40252 & $-$14.37078 \\ 
      10,020 & $-$14.55390 & $-$14.52987 & $-$14.50967 & $-$14.48493  & $-$14.46077  & $-$14.43388  & $-$14.40202 & $-$14.37081 \\ 
      10,000 & $-$14.55453 & $-$14.53047 & $-$14.51054 & $-$14.48618  & $-$14.46174  & $-$14.43522  & $-$14.40359 & $-$14.37391 \\ 
&&&\\
\hline 
\end{tabular}   
\tablefoot{The full table is available only electronically via CDS, 
a small portion is shown here for guidance on its form and content.}
\label{tab:XPf-bins}
\end{table*} 

\section{X-rays and ultraviolet observations with the Swift satellite}\label{appendix:Swift}

The {\it Swift} observations were acquired with the on-board instruments
X-Ray Telescope \citep[XRT;][]{2005SSRv..120..165B} and UltraViolet Optical
Telescope \citep[UVOT;][]{2005SSRv..120...95R}.  The XRT covers the 0.3--10
keV X-ray band, whereas UVOT data were collected mainly using UVM2 or UVW2
ultraviolet filters, with a few cases acquired in the UVW1 and U bands
\citep[see][for details]{2008MNRAS.383..627P, 2011AIPC.1358..373B}.  The two
instruments observed all objects simultaneously, with the exception of GaSS
2-51 for which no UVOT data were acquired due to instrument safety reasons
because of the close proximity of the target to the 5$^{\rm th}$ magnitude
star $\zeta$ Lib.  Exposure times for each pointing are reported in
Table~\ref{tab:Swift}.  All observations were reduced within the {\sc
ftools} environment \citep{1995ASPC...77..367B}; details on the data
analysis process can be found in Paper I.

Targets were generally observed in `snapshot' mode, i.e.  with pointings of
duration less than 1000 s.  When needed, especially for the X--ray data,
observations were stacked together to increase the signal-to-noise ratio. 
We remark that the choice of an aperture photometry radius larger than usual
(5$''$) was needed for some UVOT pointings, especially those acquired
starting with August 2023 in order to address the excess broadening of the
UVOT images point spread function produced by issues in the spacecraft
attitude control \citep{2023GCN.34633....1C, 2024MNRAS.530....1H,
2024ApJ...960L..18S}.  A check to avoid detector small scale sensitivity
issues\footnote{{\tt https://swift.gsfc.nasa.gov/analysis/
uvot\_digest/sss\_check.html}} potentially affecting the UVOT photometry was
also performed.

Photometric results in the UVOT Vega-system are reported in
Table~\ref{tab:Swift} together with those of X--ray pointings: from the
latter no emission was detected in the 0.3–10 keV range for any of the
sources; 3-$\sigma$ flux limits were determined using the procedure of
\citet{1986ApJ...303..336G} with the exception of GaSS 2-55 for which the
X--ray flux limit was evaluated through a more conservative approach using
the {\sc xspec} package v.  12.10.0 \citep{1996ASPC..101...17A} due to the
high background during the XRT observation.

To define the sensitivity to accretion processes of observations in the UVOT
bands of our program stars (cf.  Sect~\ref{UVOT} for details), we searched
the {\it Swift} archive for serendipitous observations of field, high
Galactic latitude, and non-variable red giants with accurate spectral
classifications from \citet{1999mctd.book.....H}, which are sufficiently
faint to avoid overloading the {\it Swift} detectors.  We found 16 suitable
red giants in the archive, which we supplemented with ToO observations of
four additional objects (all of M spectral type).  The results are
summarized below in Table~\ref{tab:M2W2B}.

\begin{table*}
\caption{{\it Swift} UVOT observations in the W1, M2, and W2 bands of non
variable field red giants located at high Galactic latitudes, aiming to
derive the reference level of their photospheric emission (see
Sect.~\ref{Swift} for details).}
\label{tab:M2W2B}
\begin{center}
\begin{tabular}{lllcccrc}
\hline\hline
\noalign{\smallskip}
&&&\multicolumn{3}{c}{{\it Swift} UVOT} & & GALEX\\ \cline{4-6}
\multicolumn{1}{c}{star} & SpType & 
\multicolumn{1}{c}{date} & UVW1 & UVM2 & UVW2 & \multicolumn{1}{c}{$B$} &NUV \\ 
\noalign{\smallskip}
\hline
\noalign{\smallskip}
 HD   2059    &   K0 III   & 11/05/2017    &   13.21$\pm$0.04    &    17.31$\pm$0.06     &   14.32$\pm$0.04 &    10.29  &  18.008$\pm$0.031 \\ 
 HD  88283    &   K1 III   & 16/05/2010    &   13.02$\pm$0.04    &    17.14$\pm$0.09     &   14.22$\pm$0.04 &    10.23  &  18.229$\pm$0.051 \\ 
 HD   3704    &   K1/2 III & 28/05/2017    &   12.85$\pm$0.04    &    17.11$\pm$0.11     &   13.83$\pm$0.04 &    10.10  &                   \\ 
 HD  89603    &   K2 III   & 23/11/2007    &   12.21$\pm$0.04    &    16.30$\pm$0.05     &   13.42$\pm$0.04 &     9.48  &  17.318$\pm$0.025 \\ 
 HD  72211    &   K3 III   & 26/03/2005    &   12.95$\pm$0.04    &    17.67$\pm$0.06     &   14.13$\pm$0.04 &    10.05  &  18.844$\pm$0.074 \\ 
 HD  91605    &   K3/4 III & 08/07/2010    &   12.83$\pm$0.04    &    17.46$\pm$0.10     &   14.09$\pm$0.04 &    10.01  &  18.601$\pm$0.045 \\ 
 HD  29943    &   K4 III   & 12/12/2021    &   12.93$\pm$0.04    &    17.29$\pm$0.11     &   13.98$\pm$0.04 &    10.22  &  18.656$\pm$0.046 \\ 
 HD   1208    &   M0 III   & 20/04/2016$^b$&   12.98$\pm$0.04    &    17.44$\pm$0.09     &   13.95$\pm$0.04 &    10.07  &  18.801$\pm$0.071 \\ 
 HD 184438    &   M0 III   & 01/07/2019    &   12.21$\pm$0.04    &    16.57$\pm$0.05     &   13.08$\pm$0.04 &     9.35  &                   \\ 
 HD 214003    &   M1 III   & 19/09/2016$^b$&   12.18$\pm$0.04    &    16.62$\pm$0.07     &   13.10$\pm$0.04 &     9.26  &  18.129$\pm$0.036 \\ 
 HD 126308$^a$&   M2 III   & 20/03/2023    &   13.36$\pm$0.04    &    17.72$\pm$0.15     &   14.27$\pm$0.04 &    10.37  &  19.045$\pm$0.040 \\ 
%HD 126308$^a$&   M2 III   & 23/03/2023    &   13.40$\pm$0.04    &    17.49$\pm$0.14     &   14.31$\pm$0.04 &           &                   \\ 
 HD  16756    &   M2 III   & 11/09/2013    &   12.54$\pm$0.04    &    16.97$\pm$0.04     &   13.53$\pm$0.04 &     9.56  &  18.504$\pm$0.067 \\ 
 HD   7235    &   M2/3 III & 26/01/2013    &   13.00$\pm$0.05    &    17.47$\pm$0.04     &   14.10$\pm$0.04 &    10.16  &  18.819$\pm$0.076 \\ 
 HD 139867    &   M2/3 III & 24/01/2018    &   11.93$\pm$0.04    &    16.74$\pm$0.08     &   12.73$\pm$0.04 &     9.00  &                   \\ 
 HD  31518    &   M3 III   & 14/11/2009$^b$&   11.82$\pm$0.04    &    16.13$\pm$0.05     &   12.86$\pm$0.04 &     8.84  &                   \\ 
 HD 154764$^a$&   M3 III   & 24/03/2023    &   13.77$\pm$0.04    &    18.50$\pm$0.20     &   14.78$\pm$0.04 &    10.77  &  19.651$\pm$0.119 \\ 
 HD  66424    &   M5 III   & 06/05/2019    &   12.93$\pm$0.04    &    17.45$\pm$0.08     &   13.82$\pm$0.04 &    10.00  &  18.657$\pm$0.057 \\ 
 HD  83455$^a$&   M6 III   & 24/03/2023    &   12.67$\pm$0.04    &    16.80$\pm$0.09     &   13.63$\pm$0.04 &     9.78  &  18.552$\pm$0.063 \\ 
 HD  95384$^a$&   M7/8 III & 20/03/2023    &   13.20$\pm$0.04    &    18.33$\pm$0.17     &   14.06$\pm$0.04 &    10.33  &  19.941$\pm$0.157 \\                 
\noalign{\smallskip}
\hline
\end{tabular} 
\label{tab:appendixSwift}
\tablefoot{$^a$ = stars that we observed as Target-of-Opportunity sources with {\it Swift} UVOT. 
$^b$ = stars with observations in W1 and/or W2 bands carried out on a different date, 
the quoted one being for the M2 band. $B$ band magnitude as listed by SIMBAD. GALEX NUV magnitudes
are in the native AB system.} 
\end{center}
\end{table*}  

\end{appendix}
\end{document}